\documentclass[12pt]{article}
\pdfoutput=1

\usepackage{amsmath}
\usepackage{amsfonts}
\usepackage{amscd}
\usepackage{amsthm}
\usepackage{setspace}

\usepackage{graphicx}
\usepackage{authblk}
\usepackage{caption}
\usepackage{ytableau}
\usepackage{mathtools}

\def\Z{\mathbb{Z}}

\def\R{\mathbb{R}}
\def\C{\mathbb{C}}
\def\P{\mathbb{P}}
\def\H{\mathbb{H}}

\begin{document}

\begin{titlepage}

\begin{flushright}
KEK-TH 2081 
\end{flushright}

\vskip 1cm

\begin{center}

{\large Nongeometric heterotic strings and dual F-theory \\
with enhanced gauge groups}

\vskip 1.2cm

Yusuke Kimura$^1$ 
\vskip 0.4cm
{\it $^1$KEK Theory Center, Institute of Particle and Nuclear Studies, KEK, \\ 1-1 Oho, Tsukuba, Ibaraki 305-0801, Japan}
\vskip 0.4cm
E-mail: kimurayu@post.kek.jp

\vskip 1.5cm
\abstract{\par Eight-dimensional nongeometric heterotic strings were constructed as duals of F-theory on $\Lambda^{1,1}\oplus E_8\oplus E_7$ lattice polarized K3 surfaces by Malmendier and Morrison. We study the structure of the moduli space of this construction. There are special points in this space at which the ranks of the non-Abelian gauge groups on the 7-branes in F-theory are enhanced to 18. We demonstrate that the enhanced rank-18 non-Abelian gauge groups arise as a consequence of the coincident 7-branes, which deform stable degenerations on the F-theory side. This observation suggests that the non-geometric heterotic strings include nonperturbative effects of the coincident 7-branes on the F-theory side. The gauge groups that arise at these special points in the moduli space do not allow for perturbative descriptions on the heterotic side. 
\par We also construct a family of elliptically fibered Calabi--Yau 3-folds by fibering K3 surfaces with enhanced singularities over $\P^1$. Highly enhanced gauge groups arise in F-theory compactifications on the resulting Calabi--Yau 3-folds.}  

\end{center}
\end{titlepage}

\tableofcontents
\section{Introduction}
\par F-theory \cite{Vaf, MV1, MV2} is a theoretical framework that extends type IIB superstring theory to a non-perturbative regime. The F-theory/heterotic duality \cite{Vaf, MV1, MV2, Sen, FMW} enables us to study non-perturbative aspects of heterotic strings. F-theory/heterotic duality states that heterotic string compactification \footnote{See, e.g., \cite{BHW0504, BHW0507, ACGLP1307, AGS1402, OS1402, AT1405, OHS1509, COO1605, AOMSS1806, COOS1810} for recent studies of heterotic strings.} on an elliptically fibered Calabi--Yau $n$-fold and F-theory compactification on a K3 fibered Calabi--Yau $n+1$-fold describe equivalent physical theories. This duality is strictly formulated when the stable degeneration \cite{FMW, AM} \footnote{See, for example, \cite{AHK, BKW, BKL, CGKPS, MizTan} for recent studies on the stable degenerations in F-theory/heterotic duality.} where K3 fibers split into pairs of 1/2 K3s is considered on the F-theory side. 
\par Recently, dual models of heterotic strings and F-theory, in which heterotic compactification does not allow for a geometric interpretation, have been discussed in \cite{MM}. The Narain space \cite{Narain}
\begin{equation}
D_{2,18} / O(\Lambda^{2,18})
\end{equation}
is the moduli space of eight-dimensional heterotic strings. $D_{2,18}$ denotes the symmetric space of $O(2,18)$, i.e., $D_{2,18}:=O(2)\times O(18) \backslash O(2,18)$. The double cover of this moduli space,
\begin{equation}
D_{2,18} / O^+ (\Lambda^{2,18}),
\end{equation}
coincides with the F-theory moduli on elliptically fibered K3 surfaces with a global section. ($O^+(\Lambda^{2,18})$ denotes an index two subgroup of $O(\Lambda^{2,18})$, defined as $O^+(\Lambda^{2,18}):=O(\Lambda^{2,18})\cap O^+(2,18)$.) This defines the F-theory/heterotic duality. In \cite{MM}, using the F-theory/heterotic duality Malmendier and Morrison constructed ``non-geometric'' eight-dimensional heterotic string compactifications on the 2-torus that possesses (a subgroup of) $O^+(\Lambda^{2,18})$ symmetry. The action of $O^+(\Lambda^{2,18})$ mixes the complex structure moduli, K\"ahler moduli, and the moduli of expectation values of the Wilson line. Consequently, heterotic strings that possesses $O^+(\Lambda^{2,18})$ symmetry do not yield a geometric interpretation \footnote{Non-geometric type II theories were discussed in \cite{HMW}.}. 
\par In \cite{MM}, the moduli of F-theory on elliptically fibered K3 surfaces with a section with unbroken $\mathfrak{e}_8\mathfrak{e}_7$ gauge algebra and the heterotic duals were considered for constructing non-geometric heterotic strings. Because the $\mathfrak{e}_8\mathfrak{e}_7$ algebra is unbroken for this specific case, we can take the orthogonal complement of the sum of $E_8\oplus E_7$ and $H$ ($H$ is the lattice generated by the fiber and the zero-section) inside the K3 lattice to obtain the moduli space of eight-dimensional F-theory on an elliptically fibered K3 surface with an $E_8E_7$ singularity as
\begin{equation}
D_{2,3} / O^+ (L^{2,3}).
\end{equation}
$L^{2,3}$ is the orthogonal complement of the lattice $H\oplus E_8\oplus E_7$ in the K3 lattice $\Lambda_{K3}$. The mathematical results of Kumar \cite{Kumar} and Clingher and Doran \cite{ClingherDoran2011, ClingherDoran2012} gave the Weierstrass equations of elliptically fibered K3 surfaces with a section with an $E_8E_7$ singularity, the coefficients of which are Siegel modular forms \footnote{The Siegel modular forms of genus two were discussed in the context of string compactifications in \cite{Curio1998, MMPufolds, BFMUfold}.} of even weight. The periods $\underline{\tau}$ of this K3 surface give a point in the Siegel upper half-space of genus two, $\H_2$. Using this result, Malmendier and Morrison constructed heterotic strings with $O^+(L^{2,3})$ symmetry with the variable $\underline{\tau}$. The action of $ O^+(L^{2,3})$ mixes complex structure moduli, K\"ahler moduli, and the moduli of Wilson line values. Therefore, the resulting heterotic strings are non-geometric. The Weierstrass equation of an elliptically fibered K3 surface with an $E_8E_7$ singularity determines a point in $D_{2,3}$, and the periods of this K3 surface determine the point $\underline{\tau}$ in $\H_2$. Because $D_{2,3}$ and the Siegel upper half-space of genus two, $\H_2$, are isomorphic, this construction defines a correspondence between F-theory on elliptic K3 surfaces with a section with an $E_8E_7$ singularity and non-geometric heterotic strings with $O^+(L^{2,3})$ symmetry \footnote{Connections of lattice polarized K3 surfaces, $O^+(\Lambda^{2,2})$-modular forms, and non-geometric heterotic strings were discussed in \cite{MMS}. K3 surfaces with $\Lambda^{1,1}\oplus E_7\oplus E_7$ lattice polarization and the construction of non-geometric heterotic strings were studied in \cite{CMSE7}.}. We also refer the reader to \cite{GJ1412, LMV1508, FGLMM1603, MS1609, GLMM1611, FM1708, FGLMM1712} for recent studies on non-geometric strings. 
\par In this study, we investigate the structure of the moduli space parameterizing F-theory on elliptic K3 surfaces with $E_8E_7$ singularity with a section
\begin{equation}
D_{2,3} / O^+ (L^{2,3}),
\end{equation}
to analyze the special points of the moduli space at which the ranks of the non-Abelian gauge groups on the F-theory side are enhanced to 18. Because rank-18 non-Abelian gauge groups do not embed into $E_8\times E_8\times U(1)^4$ or $SO(32)\times U(1)^4$, the non-geometric heterotic duals do not yield a perturbative interpretation of these gauge groups.
\par The divisors in a K3 surface form a lattice known as the N\'eron-Severi lattice. The rank of this lattice is referred to as the Picard number $\rho$, and for an elliptically fibered complex K3 surface with a section, the Picard number $\rho$ takes values from two to 20. The elliptic K3 surfaces with Picard number 20 are called attractive K3 surfaces. The K3 surfaces with Picard number $\rho$ in the complex structure moduli of K3 surfaces form a $(20-\rho)$-dimensional subset. In particular, this means that the moduli of attractive K3 surfaces are discrete and zero-dimensional. Attractive K3 surfaces are known to be labelled by triplets of integers \cite{PS-S, SI}. An increase in the Picard number $\rho$ implies that the rank of the gauge group increases. This equivalently means that the gauge group that arises in the F-theory compactification is enhanced. Of the Picard number $\rho=20$ for an attractive K3 surface with a section, two of these degrees of freedom are used for the zero-section and the fiber. The non-Abelian gauge group generates a lattice, and the rank of this lattice comprise at most the remaining 18 degrees of freedom. An elliptic K3 surface with a section is referred to as the {\it extremal} K3 surface when the rank of the non-Abelian gauge group is the maximum 18. An extremal K3 has Picard number 20, which is the sum of the rank 18 of the non-Abelian gauge group and the rank two of the lattice generated by the zero-section and the fiber. Therefore, such a surface is always attractive. 
\par The points in the complex structure moduli of K3 surfaces that correspond to the extremal K3s form a subset of the attractive K3 surfaces, and thus they are zero-dimensional and discrete. The heterotic strings dual to F-theory on the extremal K3 surfaces do not allow for a perturbative interpretation of the arising gauge groups. In this sense, the points that correspond to the extremal K3 surfaces in the F-theory moduli of elliptic K3 surfaces conspicuously demonstrate the non-geometric nature of dual heterotic strings.
\par In this work, we analyze the physics both on the F-theory and heterotic sides of compactifications at special points in the moduli that correspond to F-theory on extremal K3s. 
\par In particular, we study the points in the moduli with an unbroken $\mathfrak{e}_8\mathfrak{e}_7$ gauge algebra at which elliptically fibered K3 surfaces become extremal. In addition, we study an extremal K3 surface obtained as the Jacobian fibration \footnote{See \cite{Cas} for the construction of the Jacobians of elliptic curves.} of a genus-one fibered K3 surface that lacks a global section \footnote{F-theory compactifications on genus-one fibrations without a section have recently been studied, e.g., in \cite{BM, MTsection, AGGK, KMOPR, GGK, MPTW, MPTW2, BGKintfiber, CDKPP, LMTW, K, K2, ORS1604, KCY4, CGP, Kdisc, Kimura1801, AGGO1801, Kimura1806, TasilectWeigand, CLLO, TasilectCL, HT}.} constructed as a (3,2) hypersurface in $\P^2\times\P^1$, as studied in \cite{K}. Taking the Jacobian fibration of a genus-one fibered K3 surface yields an elliptic K3 surface with a global section. We analyze the physics of F-theory compactification on this extremal K3 and the heterotic dual. 
\par It was found in \cite{KRES} that some extremal K3 surfaces can be viewed as the deformations of K3 surfaces obtained as the reverse of the stable degeneration, in which a K3 surface splits into two rational elliptic surfaces. This can be understood as an effect of coincident 7-branes from the F-theory viewpoint. Under the stable degeneration limit, the dual of the F-theory model yields a geometric heterotic string. Thus, we deduce that non-geometric heterotic strings as duals of F-theory on extremal K3s are a consequence of the transitions from geometric heterotic strings when 7-branes are coincident. This suggests that non-geometric heterotic strings include the non-perturbative effects of coincident 7-branes on the F-theory side. In the heterotic language, these deformations from the stable degeneration are owing to the presence of 5-branes.
\par Furthermore, we construct elliptically fibered Calabi--Yau 3-folds with a section by fibering K3 Jacobian fibrations, as discussed in Section \ref{sssec3.2.2} over $\P^1$. We analyze F-theory compactifications on these elliptically fibered Calabi--Yau 3-folds \footnote{See, for example, \cite{GM, KT0906, KMT0911, KMT1008, MTmatter, GM2, BG1112, MT1201, MT1204, T1205, JT1605, MPT1610, MMP1711, HT1805, LLW1810} for recent studies on F-theory compactifications on elliptic Calabi--Yau 3-folds. F-theory compactifications on Calabi--Yau 3-folds with a terminal singularity are discussed in \cite{BM, AGW1612, GW1804}.}. Recent studies have mainly considered local models in F-theory model building \cite{DWmodel, BHV, BHV2, DWGUT}. However, global F-theory models must be considered to discuss gravity and issues of the early universe and inflation. In this work, we study F-theory compactifications on elliptic Calabi--Yau 3-folds from the global perspective. Highly enhanced non-Abelian gauge groups arise in the compactifications on our constructions of the elliptic Calabi--Yau 3-folds. These Calabi--Yau 3-folds are constructed as the Jacobians of genus-one fibered Calabi--Yau 3-folds without a section, as hypersurfaces in the products of projective spaces. Similar constructions of genus-one fibered Calabi--Yau 4-folds can be found in \cite{KCY4}.
\par In Section \ref{sec2}, we briefly review the construction of non-geometric heterotic strings considered in \cite{MM}. We also review F-theory and extremal K3 surfaces. In Section \ref{sec3}, we study special points of the 8-dimensional non-geometric heterotic string moduli at which the ranks of the non-Abelian gauge groups of F-theory duals reach 18. We find that eight-dimensional non-geometric heterotic strings at these points in the moduli are obtained as the transitions from geometric heterotic strings when 7-branes are coincident on the F-theory side. We also confirm that these points satisfy 5-brane solutions. Elliptic K3 surfaces generally admit distinct elliptic fibrations. Therefore, even when the complex structure of the compactification space is fixed, different gauge symmetries arise in F-theory compactification for distinct elliptic fibrations. An elliptically fibered K3 surface with a type $II^*$ fiber and a type $III^*$ fiber always admits another elliptic fibration with a type $I^*_{10}$ fiber \cite{ClingherDoran2012, MMS}. Birational maps from an elliptic fibration with a type $II^*$ fiber and a type $III^*$ fiber to an elliptic fibration with a type $I^*_{10}$ fiber were considered in \cite{MM}. These birational maps relate $E_8\times E_8$ heterotic string to $SO(32)$ heterotic string. Using these birational maps, we also study some applications to $SO(32)$ heterotic strings in Section \ref{ssec3.3}. In Section \ref{sec4}, we discuss six-dimensional F-theory compactifications on elliptic Calabi--Yau 3-folds with a section built by fibering the Jacobian K3 surface discussed in Section \ref{sssec3.2.2} over $\P^1$. We determine the Mordell--Weil groups for specific elliptic Calabi--Yau 3-folds. As a result, we determine the global structure of the gauge groups. F-theory on these specific elliptic Calabi--Yau 3-folds do not have a $U(1)$ gauge field. We also determine candidate matter spectra for some models. We deduce candidate matter spectra directly from the global defining equations of the Calabi-Yau 3-folds. We confirm that the anomaly cancels for these models. We state our concluding remarks in Section \ref{sec5}.

\section{Non-geometric heterotic strings, F-theory, and extremal K3}
\label{sec2}
\subsection{Review of F-theory}
\label{ssec2.1}
In this section, we briefly review F-theory compactification and the non-Abelian gauge groups that arise on 7-banes. 
\par F-theory is compactified on Calabi--Yau spaces that admit genus-one fibrations. The modular parameter of an elliptic fiber is identified with the axio-dilaton $C_0+ie^{-\phi}$. This F-theory formulation enables the axio-dilaton to have the monodromy property. A genus-one fibration does not necessarily have a global section: there are situations in which it admits a global section and those in which it does not. F-theory models discussed in this note admit a section \footnote{F-theory models with a global section have been intensively studied in the literature. See, for example, \cite{MorrisonPark, MPW, BGK, BMPWsection, CKP, BGK1306, CGKP, CKP1307, CKPS, AL, EKY1410, LSW, CKPT, CGKPS, MP2, BMW2017, CL2017, BMW1706, EKY1712, KimuraMizoguchi, EK1802, Kimura1802, LRW2018, EK1808, MizTani2018}.}, namely we consider F-theory models on elliptically fibered Calabi--Yau manifolds that admit a global section. (However, we consider genus-one fibered Calabi--Yau spaces lacking a global section in Sections \ref{sssec3.2.2} and \ref{sec4} to construct elliptically fibered Calabi--Yau spaces with a global section. We take the Jacobian fibrations of Calabi--Yau genus-one fibrations to yield Calabi--Yau elliptic fibrations with a section.) F-theory compactified on a complex $(n+1)$-dimensional Calabi--Yau manifold yields a $(10-2n)$-dimensional theory. Fibers degenerate along the codimension one locus in the base space. This locus is referred to as the discriminant locus, and 7-branes are wrapped on the components of this locus. The fiber type over the component on which 7-branes are wrapped determines the non-Abelian gauge group that arises on the 7-branes \cite{MV2, BIKMSV}. 
\par When an elliptic fibration admits a global section, it also admits a transformation to the Weierstrass form. In this case, the type of a singular fiber can be determined from the vanishing orders of the Weierstrass coefficients. The types of the singular fibers of an elliptic surface were classified, and their monodromies and j-invariants were computed, by Kodaira in \cite{Kod1, Kod2}. Techniques to determine the types of the singular fibers of an elliptic fibration are described in \cite{Ner, Tate}. The types of singular fibers of an elliptic surface, their monodromies and j-invariants, the vanishing orders of the Weierstrass coefficients, and the corresponding singularity types are listed in Tables \ref{tabfibertypes in 2.1} and \ref{tabvanishingorder in 2.1}. The set of sections of an elliptic fibration forms a group, which is called the Mordell--Weil group. The rank of the Mordell--Weil group is equal to the number of $U(1)$ gauge fields that arise in F-theory compactification. 

\begingroup
\renewcommand{\arraystretch}{1.1}
\begin{table}[htb]
  \begin{tabular}{|c|c|r|c|c|} \hline
fiber type & j-invariant & monodromy  & order of monodromy & singularity type \\ \hline
$I^*_0$ & regular & $-\begin{pmatrix}
1 & 0 \\
0 & 1 \\
\end{pmatrix}$ & 2 & $D_4$\\ \hline
$I_b$ & $\infty$ & $\begin{pmatrix}
1 & b \\
0 & 1 \\
\end{pmatrix}$ & infinite & $A_{b-1}$\\
$I^*_b$ & $\infty$ & $-\begin{pmatrix}
1 & b \\
0 & 1 \\
\end{pmatrix}$ & infinite & $D_{b+4}$\\ \hline
$II$ & 0 & $\begin{pmatrix}
1 & 1 \\
-1 & 0 \\
\end{pmatrix}$ & 6 & none.\\
$II^*$ & 0 & $\begin{pmatrix}
0 & -1 \\
1 & 1 \\
\end{pmatrix}$ & 6 & $E_8$\\ \hline
$III$ & 1728 & $\begin{pmatrix}
0 & 1 \\
-1 & 0 \\
\end{pmatrix}$ & 4 & $A_1$\\
$III^*$ & 1728 & $\begin{pmatrix}
0 & -1 \\
1 & 0 \\
\end{pmatrix}$ & 4 & $E_7$\\ \hline
$IV$ & 0 & $\begin{pmatrix}
0 & 1 \\
-1 & -1 \\
\end{pmatrix}$ & 3 & $A_2$\\
$IV^*$ & 0 & $\begin{pmatrix}
-1 & -1 \\
1 & 0 \\
\end{pmatrix}$ & 3 & $E_6$\\ \hline
\end{tabular}
\caption{\label{tabfibertypes in 2.1}J-invariants, monodromies, and the corresponding singularity types of fiber types. ``Regular'' for type $I_0^*$ means that the j-invariant of this fiber type can take any finite value in $\C$.}
\end{table}
\endgroup

\begingroup
\renewcommand{\arraystretch}{1.5}
\begin{table}[htb]
\begin{center}
  \begin{tabular}{|c|c|c|c|} \hline
fiber type & ord($f$) & ord($g$) & ord($\Delta$) \\ \hline
$I_0 $ & $\ge 0$ & $\ge 0$ & 0 \\ \hline
$I_n $  ($n\ge 1$) & 0 & 0 & $n$ \\ \hline
$II $ & $\ge 1$ & 1 & 2 \\ \hline
$III $ & 1 & $\ge 2$ & 3 \\ \hline
$IV $ & $\ge 2$ & 2 & 4 \\ \hline
$I^*_0$ & $\ge 2$ & 3 & 6 \\ \cline{2-3}
 & 2 & $\ge 3$ &  \\ \hline
$I_m^*$  ($m \ge 1$) & 2 & 3 & $6+m$ \\ \hline
$IV^*$ & $\ge 3$ & 4 & 8 \\ \hline
$III^*$ & 3 & $\ge 5$ & 9 \\ \hline
$II^*$ & $\ge 4$ & 5 & 10 \\ \hline   
\end{tabular}
\caption{\label{tabvanishingorder in 2.1}The vanishing orders of the Weierstrass coefficients $f,g$ of $y^2=x^3+f\, x+g$, and the vanishing order of the discriminant $\Delta$, for the various fiber types.}
\end{center}
\end{table}  
\endgroup 

\par We also briefly review the geometries of elliptic K3 surfaces, because this is relevant to contents of Section \ref{sec3}. The second integral cohomology group $H^2(S, \Z)$ of a K3 surface $S$ is called the K3 lattice, $\Lambda_{K3}$, and this lattice contains information on the geometry of a K3 surface, such as its divisors and their intersections. A K3 lattice is a rank-22 unimodular integral lattice with signature (3,19). From the structure theorem for an indefinite integral unimodular even lattice \cite{Mil}, the following isomorphism holds:
\begin{equation}
\Lambda_{K3} \cong H^3 \oplus E_8^2,
\end{equation}
where $H$ in this isomorphism denotes the ``hyperbolic plane,'' which is a rank-two integral lattice with the following intersection form:
\begin{equation}
\begin{pmatrix}
0 & 1 \\
1 & 0 \\
\end{pmatrix}.
\end{equation}
\par A sublattice of the K3 lattice $\Lambda_{K3}$, called the N\'eron-Severi lattice $NS(S)$, determines the information on the divisors of a K3 surface $S$. The intersection form of the N\'eron-Severi lattice $NS(S)$ corresponds to the intersections of the divisors. The rank of the N\'eron-Severi lattice $NS(S)$ is called the Picard number $\rho(S)$ of the K3 surface. This number varies depending on the complex structure of the K3 surface $S$. When a K3 surface is elliptically fibered and admits a global section, the fiber $F$ and the zero-section generate the hyperbolic plane $H$. Thus, the property $H\subset NS(S)$ that the N\'eron-Severi lattice $NS(S)$ of a K3 surface $S$ contains the hyperbolic plane $H$ is equivalent \cite{Kondoauto} to the statement that the K3 surface $S$ admits an elliptic fibration with a section. 
\par When a genus-one fibered K3 surface has a section, inside the N\'eron-Severi lattice $NS(S)$, the orthogonal complement of the hyperbolic plane generated by the fiber and a section contains the information on the types of the singular fibers. A singular fiber is either $\P^1$ with a single singularity, or a sum of several smooth $\P^1$'s intersecting in a specific manner. The manner in which the $\P^1$'s intersect determines the type of a singular fiber \cite{Kod1, Kod2}. Each $\P^1$ component in a singular fiber is a divisor of the K3 surface $S$. Therefore, (except for $\P^1$'s that intersect with the zero-section) $\P^1$ components of a singular fiber belong to the orthogonal complement of the hyperbolic plane inside the K3 lattice $\Lambda_{K3}$. For this reason, the orthogonal complement captures the intersections of $\P^1$ components of singular fibers. 
\par $D_{2,18}$ modded out by (a subgroup of) the automorphism group of the orthogonal complement of the hyperbolic plane $H$ \footnote{The assumption that an elliptic K3 surface admits a global section cannot be dropped here. When a genus-one fibered K3 surface only has a multisection, the number of intersections between the multisection and the fiber is greater than one. Therefore, the multisection and the fiber do not generate the hyperbolic plane $H$.} inside the K3 lattice $\Lambda_{K3}$,
\begin{equation}
\label{autom K3 in 2.1}
D_{2,18} / O^+ (\Lambda^{2,18})
\end{equation}
yields the parameter space for the elliptic K3 surfaces with a global section. Here, $\Lambda_{2,18}$ denotes the orthogonal complement of the hyperbolic plane $H$ inside $\Lambda_{K3}$:
\begin{equation}
\Lambda_{2,18}:=H^{\perp}\subset \Lambda_{K3}.
\end{equation}
$\Lambda_{2,18}$ is an even integral unimodular lattice of signature (2,18), and this is unique up to isometry. 

\subsection{Review of non-geometric heterotic strings in Malmendier and Morrison}
\label{ssec2.2}
The moduli space of eight-dimensional heterotic strings compactified on the 2-torus $T^2$ is given by the Narain space
\begin{equation}
D_{2,18}/O(\Lambda^{2,18}).
\end{equation}
In a suitable limit, this moduli space decomposes into the product of the complex structure moduli, K\"ahler moduli, and the moduli of the Wilson line expectation values \cite{NSWheterotic}. This decomposition is preserved under the action of a parabolic subgroup $\Gamma$, but the action of $O(\Lambda^{2,18})$ does not preserve this decomposition. The action of the index-two subgroup $O^+(\Lambda^{2,18})$ still mixes the complex structure moduli, K\"ahler moduli, and the moduli of the Wilson line expectation values. Because of this, heterotic strings with $O^+(\Lambda^{2,18})$-symmetry do not yield a geometric interpretation. In this sense, they are non-geometric. This is the construction of the eight-dimensional non-geometric heterotic strings with the $O^+(\Lambda^{2,18})$-symmetry in \cite{MM}. 
\par The double cover (\ref{autom K3 in 2.1}) of the Narain space gives the moduli space of F-theory on elliptic K3 surfaces with a section. The moduli space that parameterizes F-theory on elliptic K3 surfaces with the $E_8E_7$ singularity is
\begin{equation}
D_{2,3}/ O^+(L^{2,3}),
\end{equation}
where $L^{2,3}$ is the orthogonal complement of the lattice $H\oplus E_8\oplus E_7$ in the K3 lattice $\Lambda_{K3}$ (or equivalently, the orthogonal complement of the lattice $E_8\oplus E_7$ in $\Lambda^{2,18}$). 
\par An explicit construction of non-geometric heterotic strings with $O^+(L^{2,3})$-symmetry was discussed in \cite{MM}. Utilizing the F-theory/heterotic duality, the authors constructed such non-geometric heterotic strings with an unbroken $\mathfrak{e}_8\mathfrak{e}_7$ gauge algebra as duals of F-theory on elliptic K3 surfaces with the $E_8E_7$ singularity. Their construction proceeds as follows. There is a homomorphism from the Siegel modular group $Sp_4(\R)$ to $O^+(2,3)$, and this induces an isomorphism of the Siegel upper half-space of genus two, $\H_2$, and $D_{2,3}$:
\begin{equation}
\label{isom H2 in 2.2}
\H_2\cong D_{2,3}.
\end{equation}
Under this isomorphism, the ring of $O^+(L^{2,3})$-modular forms corresponds to the Siegel modular forms with genus two of even weight \cite{Vinberg}. The Siegel modular forms with genus two of even weight form a polynomial ring generated by four generators, $\psi_4$, $\psi_6$, $\chi_{10}$, and $\chi_{12}$, of degrees 4, 6, 10, and 12, respectively \cite{Igusa}. 
\par The correspondence between the periods of elliptic K3 surfaces with a section with singularity type $E_8E_7$ and the points in the Siegel upper half-space of genus two $\H_2$ was obtained by Kumar \cite{Kumar} and Clingher and Doran \cite{ClingherDoran2011, ClingherDoran2012}. The Weierstrass coefficients of such elliptic K3 surfaces were given in terms of Siegel modular forms with genus two of even weight in their work. 
\par The Weierstrass equation of an elliptic K3 surface with a section and including the singularity type $E_8E_7$ is given by
\begin{equation}
\label{Weierstrass E8E7 in 2.2}
y^2=x^3+(a\, t^4+c\,t^3)x+t^7+b\,t^6+d\,t^5.
\end{equation}
Then (up to some scale factors), the coefficients $a,b,c,d$ are given in terms of Igusa's generators as \cite{Kumar, ClingherDoran2011, ClingherDoran2012}:
\begin{equation}
\label{coeff identification in 2.2}
a=-\frac{1}{48}\, \psi_4(\underline{\tau}), \hspace{5mm} b=-\frac{1}{864}\, \psi_6(\underline{\tau}), \hspace{5mm} c=-4\, \chi_{10}(\underline{\tau}), \hspace{5mm} d=\chi_{12}(\underline{\tau}).
\end{equation}
The elliptic K3 surface determines a point in $D_{2,3}$, and this also determines a point in $\H_2$ under the isomorphism (\ref{isom H2 in 2.2}), which is denoted by $\underline{\tau}$.
\par Now, consider a manifold $M$ with a line bundle $\cal{L}$, and choose sections $a,b,c,d$ of ${\cal L}^{{\otimes}4}$, ${\cal L}^{{\otimes}6}$, ${\cal L}^{{\otimes}10}$, and ${\cal L}^{{\otimes}12}$, respectively. When the sections $a,b,c,d$ are identified as (\ref{coeff identification in 2.2}), the compactification on $M$ (which is the 2-torus $T^2$ when an eight-dimensional heterotic string is considered) yields a heterotic string with $O^+(L^{2,3})$-symmetry. This represents the construction of non-geometric strings discussed in \cite{MM}.
\vspace{5mm}

\par In \cite{MM}, the authors determined the locations of the 5-branes in the moduli of eight-dimensional non-geometric heterotic strings by computing the locus in the dual F-theory moduli on elliptic K3 surfaces with a section with the singularity type $E_8E_7$, at which the singularity type is enhanced. The 5-brane solution is obtained from the conditions that either $c=0$ (the type $III^*$ fiber at $t=0$ is enhanced to a type $II^*$ fiber when this is satisfied) or the quartic polynomial in the discriminant $\Delta$ of the Weierstrass equation (\ref{Weierstrass E8E7 in 2.2}) degenerates, where the discriminant is
\begin{equation}
\Delta=t^9\, s^9\, (4s^2\, (at+cs)^3+27t\, (t^2+bts+ds^2)^2).
\end{equation}
The resulting 5-brane solution is \cite{MM}
\begin{equation}
\label{condition cq in 2.2}
c=0 \hspace{1cm} {\rm or} \hspace{1cm} q=0,
\end{equation}
where $q$ is a polynomial in the Weierstrass coefficients $a,b,c,d$, defined as \cite{MM}
\begin{equation}
\begin{split}
q  := & 11664\, d^5 -5832\, b^2d^4 +864\, a^3d^4+16\, a^6d^3 \\
 & +216\, a^3b^2d^3+16200\, ac^2d^3-2592\, a^2bcd^3+729\, b^4d^3+888\, a^4c^2d^2 \\
 & -5670\, ab^2c^2d^2-13500\, bc^3d^2+216\, a^4b^2c^2d+16\, a^7c^2d-3420\, a^3bc^3d \\
 & +4125\, a^2c^4d+729\, ab^4c^2d+6075\, b^3c^3d-16\, a^6bc^3+16\, a^5c^4\\
 & -216\, a^3b^3c^3+2700\, a^2b^2c^4-5625\, abc^5-729\, b^5c^3+3125\, c^6.
\end{split}
\end{equation}
\par In Section \ref{ssec3.2}, we study points in the moduli of elliptic K3 surfaces with a section with the singularity type $E_8E_7$ at which the singularity type is enhanced to rank 18. We confirm that either of the above conditions (\ref{condition cq in 2.2}) are satisfied for non-geometric heterotic duals at such points.

\subsection{Extremal K3 surfaces and enhanced gauge groups on the F-theory side}
\label{ssec2.3}
The Picard number $\rho$ of a complex elliptic K3 surface with a section takes a value between two and 20, and when the K3 surface takes the maximum value 20 it is called an attractive K3 surface \footnote{We follow the convention for the term utilized in \cite{M}, referring to a K3 surface with the Picard number 20 as an attractive K3 surface.}. The types of the singular fibers of an elliptic K3 surface correspond to the singularity type of the surface. Because the $\P^1$ components of the singular fibers embed into the N\'eron-Severi lattice $NS$, the higher the rank of the singularity type of a K3 surface, the larger the Picard number $\rho$. The N\'eron-Severi lattice $NS$ has rank 20 for an attractive K3 surface, with two of these degrees of freedom used for the fiber and the zero-section. The $\P^1$ components of the singular fibers not intersecting the zero-section are orthogonal to the fiber and the zero-section. Therefore, the rank of the singularity type that corresponds to the diagram of the intersections of these $\P^1$ components is at most 18. When the singularity type of an attractive K3 surface attains the bound of 18, it is called an {\it extremal} K3 surface \footnote{This condition is equivalent to an attractive K3 surface with a section having the Mordell--Weil rank zero.}. Physically, this means that a non-Abelian gauge group in an eight-dimensional F-theory compactification on an elliptic K3 surface that forms on the 7-branes has rank at most 18, and the rank is 18 when the K3 surface is extremal. 
\par In Section \ref{sec3}, we analyze points in the moduli of K3 surfaces (with some unbroken gauge algebras) at which K3 surfaces become extremal. As will be discussed in Section \ref{ssec3.2}, this enhancement of the gauge group can be seen as a consequence of coincident 7-branes.
\par It is known that the complex structure of an attractive K3 surface $S$ is determined by its ``transcendental lattice'' $T(S)$, which is defined as the orthogonal complement of the N\'eron--Severi lattice $NS(S)$ inside the K3 lattice $\Lambda_{K3}$ \cite{PS-S, SI}. For an attractive K3 surface, the transcendental lattice is a rank-two positive-definite integral even lattice. Under the action of $GL_2(\Z)$, the intersection form of the transcendental lattice of an attractive K3 surface is as follows:
\begin{equation}
\label{intersection abc in 2.3}
\begin{pmatrix}
2a & b \\
b & 2c \\
\end{pmatrix},
\end{equation}
where $a,b,c\in\Z$ are such that 
\begin{equation}
a \ge c \ge b \ge 0.
\end{equation}
The relation between attractive K3 surfaces and the transcendental lattices is reviewed in the context of F-theory compactification in \cite{K}. Based on this property of attractive K3 surfaces, when the intersection form of an attractive K3 surface is given by (\ref{intersection abc in 2.3}), we denote this attractive K3 surface by $S_{[2a \hspace{1mm} b \hspace{1mm} 2c]}$ in this paper. 
\par All the types of the elliptic fibrations of an attractive K3 surface with the smallest discriminant three, $S_{[2 \hspace{1mm} 1 \hspace{1mm} 2]}$, and an attractive K3 with discriminant four, $S_{[2 \hspace{1mm} 0 \hspace{1mm} 2]}$, are determined \cite{Nish} \footnote{While in general a genus-one fibered K3 surface admits both elliptic fibrations with a section and genus-one fibrations without a section, these attractive K3 surfaces only have elliptic fibrations with a global section, as proved in \cite{Keum}. F-theory compactifications on the attractive K3 surfaces $S_{[2 \hspace{1mm} 1 \hspace{1mm} 2]}$ and $S_{[2 \hspace{1mm} 0 \hspace{1mm} 2]}$ were discussed in relation to the appearances of $U(1)$ gauge fields in \cite{KimuraMizoguchi}.}. Some types of these elliptic fibrations belong to extremal K3 surfaces in the moduli of elliptic K3 surfaces with a section, as discussed in Section \ref{ssec3.2}. Because some elliptic fibrations of these attractive K3 surfaces are relevant to results in this study, we summarize the elliptic fibrations of $S_{[2 \hspace{1mm} 1 \hspace{1mm} 2]}$ and $S_{[2 \hspace{1mm} 0 \hspace{1mm} 2]}$ in Appendix \ref{secA}.

\section{Special points in eight-dimensional non-geometric heterotic moduli and F-theory duals}
\label{sec3}
\subsection{Summary of discussion}
\label{ssec3.1}
\par Under the stable degeneration limit, an elliptic K3 surface splits into a pair of 1/2 K3s, i.e., a pair of rational elliptic surfaces. F-theory/heterotic duality is precisely formulated when this limit is taken. The Picard number of a rational elliptic surface is always 10, and the rank of the singularity type ranges from zero to eight. A rational elliptic surface with the singularity type of rank eight is called an extremal rational elliptic surface. Some extremal rational elliptic surfaces appear in Section \ref{ssec3.2}. The fiber types of the extremal rational elliptic surfaces were classified in \cite{MP}. We list the types of singular fibers for the extremal rational elliptic surfaces in Table \ref{tabextremalRES} in Appendix \ref{secB}. The complex structure of an extremal rational elliptic surface is uniquely determined by the types of the singular fibers, except for those surfaces with two type $I_0^*$ fibers \cite{MP}. The complex structures of the extremal rational elliptic surfaces with two type $I_0^*$ fibers depend on the j-invariant $j$ of the elliptic fibers. The fixed $j$ determines the complex structure of an extremal rational elliptic surface with two type $I_0^*$ fibers. Given these, we denote, for example, the extremal rational elliptic surfaces with singular fibers of type $II^*$ and $II$ as $X_{[II, \hspace{1mm} II^*]}$. Here, $n$ is simply employed to denote the type $I_n$ fiber, and $m^*$ represents the type $I_m^*$ fiber. The extremal rational elliptic surface with type $II^*, I_1, I_1$ fibers is denoted by $X_{[II^*, \hspace{1mm} 1, 1]}$. An extremal rational elliptic surface with two type $I_0^*$ fibers is denoted by $X_{[0^*, \hspace{1mm} 0^*]}(j)$, because this surface depends on the j-invariant $j$ of an elliptic fiber. The same notational convention is adopted in \cite{KRES}. As discussed in \cite{KRES}, the singular fibers of an elliptic K3 surface obtained as the reverse of the stable degeneration is the sum of those of a pair of rational elliptic surfaces. Thus, the rank of the non-Abelian gauge group formed in F-theory compactification on an elliptic K3 surface obtained as the reverse of the stable degeneration can be up to 16. The heterotic dual of this compactification allows for a geometric interpretation by taking the large radius limit. 
\par However, at some special points in the moduli of elliptically fibered K3 surfaces, K3 surfaces become extremal, and the non-Abelian gauge groups forming in F-theory compactifications on such extremal K3 surfaces have rank 18. Through dimensional reduction of 10-dimensional heterotic string to eight dimensions, $U(1)^2$ groups arise from each of the metric $g_{MN}$ and the $B$ field $B_{MN}$. Therefore, perturbative heterotic strings can have gauge groups up to $E_8\times E_8\times U(1)^4$ (or $SO(32)\times U(1)^4$). Rank 18 non-Abelian gauge groups, such as $E_8\times E_8\times SU(3)$, do not allow for perturbative descriptions on the heterotic side.
\par There is a special deformation of the stable degeneration in which 7-branes coincide and singular fibers collide, as studied in \cite{KRES}. In this situation, the type of a singular fiber of the resulting K3 surface is enhanced, and the rank of the singularity type increases. Consequently, some extremal K3 surfaces are obtained in this special deformation of (the reverse of) the stable degeneration, as discussed in \cite{KRES}. 
\par F-theory models on some extremal K3 surfaces are dual to non-geometric heterotic strings constructed by Malmendier and Morrison \cite{MM}, as will be discussed in Section \ref{ssec3.2}. We demonstrate in Section \ref{ssec3.2} that these F-theory models on extremal K3s that are dual to non-geometric strings constructed in \cite{MM} can be obtained as special deformations of F-theory models in the stable degeneration limit, as a consequence of coincident 7-branes. Because of this, the duality between eight-dimensional F-theory models with rank 18 gauge groups and non-geometric heterotic strings appears to suggest that non-geometric strings in \cite{MM} include the non-perturbative effect in which 7-branes coincide on the F-theory side. In addition, we confirm that the non-geometric heterotic duals of these F-theory models on extremal K3s satisfy 5-brane solutions (\ref{condition cq in 2.2}). The non-geometric heterotic duals of eight-dimensional F-theory models with rank 18 gauge groups satisfy multiple 5-brane solutions. Therefore, on the heterotic side, these deformations from the stable degeneration can be understood as a consequence of the insertion of multiple 5-branes. We also discuss some applications to $SO(32)$ heterotic strings in Section \ref{ssec3.3}.

\subsection{F-theory on extremal K3s and non-geometric heterotic duals}
\label{ssec3.2}
We discuss special points in the moduli of eight-dimensional non-geometric heterotic strings, at which the non-Abelian gauge groups that form on the F-theory side have rank 18. Eight-dimensional heterotic strings are compactified on the 2-torus $T^2$, and the dual F-theory models are compactified on elliptic K3 surfaces.
\par In particular, we consider $E_8\times E_8$ non-geometric heterotic strings with an unbroken $\mathfrak{e}_8\mathfrak{e}_7$ gauge algebra, as constructed in \cite{MM}. We also discuss non-geometric heterotic strings as duals of F-theory on the extremal K3 surface with the $E_6^3$ singularity. It is known \cite{MorrisonK3} that an attractive K3 surface always admits a Shioda--Inose structure \cite{SI, Inose}, namely, it admits an elliptic fibration for which the singular fibers include two type $II^*$ fibers. It follows from this fact that the extremal K3 surface with the $E_6^3$ singularity admits a birational transformation to some elliptic fibration with two type $II^*$ fibers. (In fact, from Table 2 in \cite{SZ} we observe that the extremal K3 surface with the $E_6^3$ singularity is $S_{[2 \hspace{1mm} 1 \hspace{1mm} 2]}$, and all the elliptic fibrations of this are known \cite{Nish}. These fibrations are listed in Table \ref{tab[212]fibration} in Appendix \ref{secA}. Fibration no. 5 corresponds to the $E_6^3$ singularity, and a certain birational transformation maps this fibration to fibration no. 1.) The singularity type of an attractive K3 surface with two type $II^*$ fibers includes $E_8^2$, and thus the attractive K3 surface with a Shioda--Inose structure to which the extremal K3 surface with the $E_6^3$ singularity is mapped belongs to the moduli with an unbroken $\mathfrak{e}_8\mathfrak{e}_7$ gauge algebra. From this observation, we deduce that the Weierstrass coefficients of the extremal K3 surface with the $E_6^3$ singularity are given by Siegel modular forms. This observation implies that F-theory on the extremal K3 surface with the $E_6^3$ singularity is related to some non-geometric heterotic dual. 
\par We determine the complex structures of these extremal K3 surfaces, and we find that some deformations from elliptic K3 surfaces, obtained as sums of pairs of identical 1/2 K3s as the reverse of the stable degenerations, yield these extremal K3 surfaces. These deformations can be understood as the effects of coincident 7-branes.

\subsubsection{Unbroken $\mathfrak{e}_8\mathfrak{e}_7$ gauge algebra}
\label{sssec3.2.1}
We discuss the points in the moduli of elliptic K3 surfaces with a global section with the singularity type $E_8E_7$ at which K3 surfaces become extremal, namely those points at which the singularity types have rank 18. 
\par A generic elliptic K3 surface in this moduli space has the singularity type $E_8E_7$, with singular fibers of types $II^*$ and $III^*$. Identifying the Weierstrass coefficients of these K3 surfaces with sections of line bundles on the 2-torus $T^2$, as in (\ref{coeff identification in 2.2}), yields non-geometric heterotic duals, as reviewed in Section \ref{ssec2.2}. The singularity type is enhanced at special points in the moduli. 
\par The singularity types of extremal K3 surfaces were classified in \cite{SZ}. According to the result, there are four singularity types worse than $E_8E_7$ that can occur for extremal K3 surfaces. These are as follows:
\begin{equation}
E_8E_7A_2A_1, \hspace{5mm} E_8E_7A_3, \hspace{5mm} E_8^2A_2, \hspace{5mm} E_8^2A_1^2.
\end{equation}
The complex structure of an extremal K3 surface for each of these singularity types is uniquely determined \cite{SZ}. The extremal K3 surfaces with these singularity types have transcendental lattices that have the following intersection forms:
\begin{equation}
\begin{pmatrix}
6 & 0 \\
0 & 2 \\
\end{pmatrix}, \hspace{5mm} \begin{pmatrix}
4 & 0 \\
0 & 2 \\
\end{pmatrix}, \hspace{5mm} \begin{pmatrix}
2 & 1 \\
1 & 2 \\
\end{pmatrix}, \hspace{5mm} \begin{pmatrix}
2 & 0 \\
0 & 2 \\
\end{pmatrix},
\end{equation}
respectively. Therefore, we denote these extremal K3 surfaces as $S_{[6 \hspace{1mm} 0 \hspace{1mm} 2]}$, $S_{[4 \hspace{1mm} 0 \hspace{1mm} 2]}$, $S_{[2 \hspace{1mm} 1 \hspace{1mm} 2]}$, $S_{[2 \hspace{1mm} 0 \hspace{1mm} 2]}$, respectively. These extremal K3 surfaces belong to the moduli of elliptic K3 surfaces with the singularity type $E_8E_7$.
\par We discuss the Weierstrass equations of the four extremal K3 surfaces.
\par An attractive K3 surface $S_{[6 \hspace{1mm} 0 \hspace{1mm} 2]}$ \footnote{The types of elliptic fibrations of the attractive K3 surface $S_{[6 \hspace{1mm} 0 \hspace{1mm} 2]}$ were studied in \cite{BGHLMSW}.} admits an extremal fibration with the singularity type $E_8E_7A_2A_1$ \cite{SZ}. After some computation, we can determine the Weierstrass equation of this fibration. As reviewed in Section \ref{ssec2.2}, the Weierstrass equation with $E_8$ and $E_7$ singularities takes the form (\ref{Weierstrass E8E7 in 2.2}). We rewrite (\ref{Weierstrass E8E7 in 2.2}) in the homogeneous form as follows:
\begin{equation}
\label{homogeneous E8E7 in 3.2.1}
y^2=x^3+s^4\, (a\, t^4+c\,t^3s)x+t^5s^5\, (t^2+b\,ts+d\,s^2),
\end{equation}
where $[t:s]$ is the homogeneous coordinate of the base $\P^1$. In the form (\ref{homogeneous E8E7 in 3.2.1}), the type $II^*$ fiber is at $[t:s]=[1:0]$ and the type $III^*$ fiber is at $[t:s]=[0:1]$. 
\par Using an automorphism of the base $\P^1$, we may assume that the type $I_3$ fiber is at $[t:s]=[-1:1]$. Furthermore, we require that this fibration also has an $A_1$ singularity. These two conditions impose the following equations to be satisfied:
\begin{eqnarray}
\label{condition for [602] in 3.2.1}
4a^3+27b^2+54d & = & 27\, (b-\frac{3}{2})^2+162\, (b-\frac{3}{2})+81 \\ \nonumber
12a^2c+54bd & = & 81\, (b-\frac{3}{2})^2+162\, (b-\frac{3}{2})+27 \\ \nonumber
27d^2+12ac^2 & = & 81\, (b-\frac{3}{2})^2+54\, (b-\frac{3}{2}) \\ \nonumber
4c^3 & = & 27\, (b-\frac{3}{2})^2.
\end{eqnarray}
The solution to these equations (\ref{condition for [602] in 3.2.1}) is 
\begin{equation}
\label{abcd for [602] in 3.2.1}
a=-\frac{15}{2\sqrt[3]{2}}, \hspace{5mm} b=-\frac{5}{2}, \hspace{5mm} c=3\cdot \sqrt[3]{4}, \hspace{5mm} d=10.
\end{equation}
This solution yields an elliptic fibration 
\begin{equation}
\label{fibration [602] in 3.2.1}
y^2=x^3+\sqrt[3]{4}\, t^3s^4 (-\frac{15}{4}t+3s)\, x+t^5s^5(t^2-\frac{5}{2}ts+10s^2).
\end{equation}
The discriminant of this Weierstrass form is 
\begin{equation}
\label{discriminant [602] in 3.2.1}
\Delta \sim s^{10} t^9 (t-4s)^2(s+t)^3.
\end{equation}
This confirms that the elliptic fibration (\ref{fibration [602] in 3.2.1}) has a type $II^*$ fiber at $[t:s]=[1:0]$, a type $III^*$ fiber at $[t:s]=[0:1]$, a type $I_3$ fiber at $[t:s]=[-1:1]$, and a type $I_2$ fiber at $[t:s]=[4:1]$. Thus, the corresponding singularity type is $E_8E_7A_2A_1$, and according to \cite{SZ} an attractive K3 surface with this singularity type is unique, being $S_{[6 \hspace{1mm} 0 \hspace{1mm} 2]}$. We conclude from this that the Weierstrass equation (\ref{fibration [602] in 3.2.1}) gives the desired extremal fibration of $S_{[6 \hspace{1mm} 0 \hspace{1mm} 2]}$. We confirm that the solution (\ref{abcd for [602] in 3.2.1}) in fact satisfies the 5-brane solution $q=0$ in (\ref{condition cq in 2.2}). 
\par The F-theory compactification on the K3 extremal fibration (\ref{fibration [602] in 3.2.1}) times a K3 surface yields a four-dimensional theory with $N=2$ supersymmetry, and the anomaly cancellation condition requires that the total number of the 7-branes in this compactification is 24 \cite{K}. The extremal fibration (\ref{fibration [602] in 3.2.1}) with $II^*$, $III^*$, $I_3$, and $I_2$ fibers satisfies this requirement, and this provides a non-trivial consistency check.
\par According to Table 2 in \cite{SZ} (No. 324), the extremal fibration (\ref{fibration [602] in 3.2.1}) has the trivial Mordell--Wei group. Therefore, the global structure of the gauge group that forms in the F-theory compactification on the extremal fibration (\ref{fibration [602] in 3.2.1}) is 
\begin{equation}
E_8 \times E_7 \times SU(3) \times SU(2).
\end{equation}
\par Attractive K3 surface $S_{[4 \hspace{1mm} 0 \hspace{1mm} 2]}$ has an extremal fibration with the singularity type $E_8E_7A_3$. The elliptic fibrations with a section of the attractive K3 surface $S_{[4 \hspace{1mm} 0 \hspace{1mm} 2]}$ were obtained in \cite{BLe}. The extremal fibration with the singularity type $E_8E_7A_3$ is particularly given by the following (general) Weierstrass form \cite{BLe}:
\begin{equation}
y^2-2t\, xy-2t^2(t-1)\, y=x^3+t^4(t-1)^3.
\end{equation}
A translation in $y$ and completing the cube in $x$ yields the following Weierstrass form:
\begin{equation}
\label{fibration [402] in 3.2.1}
y^2=x^3+t^3s^4\, (\frac{5}{3}t-2s)\, x+t^5s^5\, (t^2-\frac{70}{27}ts+\frac{5}{3}s^2)
\end{equation}
given in the homogeneous form. The discriminant is 
\begin{equation}
\label{discriminant [402] in 3.2.1}
\Delta \sim t^9 s^{10} (t-s)^4 (27t-32s).
\end{equation}
Thus, we confirm that the Weierstrass form (\ref{fibration [402] in 3.2.1}) has a type $III^*$ fiber at $[t:s]=[0:1]$, a type $II^*$ fiber at $[t:s]=[1:0]$, a type $I_4$ fiber at $[t:s]=[1:1]$, and a type $I_1$ fiber at $[t:s]=[32:27]$. 
\par We confirm that 
\begin{equation}
a=\frac{5}{3}, \hspace{5mm} b=-\frac{70}{27}, \hspace{5mm} c=-2, \hspace{5mm} d=\frac{5}{3}
\end{equation}
satisfy 5-brane solution $q=0$ in (\ref{condition cq in 2.2}). 
\par According to Table 2 in \cite{SZ} (No. 325), the extremal fibration (\ref{fibration [402] in 3.2.1}) has the trivial Mordell--Wei group. Therefore, the global structure of the gauge group that forms in F-theory compactification on the extremal fibration (\ref{fibration [402] in 3.2.1}) is 
\begin{equation}
E_8\times E_7 \times SU(4).
\end{equation}
\par The attractive K3 surface $S_{[2 \hspace{1mm} 1 \hspace{1mm} 2]}$ has an extremal fibration with the singularity type $E_8^2A_2$. The Weierstrass equation of this extremal fibration is given as follows \cite{Shioda2008}:
\begin{equation}
\label{fibration another [212] in 3.2.1}
y^2=x^3+t^5\, (t-\frac{1}{4})^2,
\end{equation}
or in the homogeneous form, 
\begin{equation}
\label{fibration [212] in 3.2.1}
y^2=x^3+t^5s^5\, (t^2-\frac{1}{2}\, ts+\frac{1}{16}s^2).
\end{equation}
The discriminant is given by
\begin{equation}
\label{discriminant [212] in 3.2.1}
\Delta \sim t^{10}s^{10}\, (t-\frac{1}{4}s)^4.
\end{equation}
It can be confirmed from (\ref{fibration [212] in 3.2.1}) and (\ref{discriminant [212] in 3.2.1}) that there are two type $II^*$ fibers at $[t:s]=[0:1]$ and $[t:s]=[1:0]$, and there is a type $IV$ fiber at $[t:s]=[1:4]$. 
\par We find that 
\begin{equation}
\label{abcd for [212] in 3.2.1}
a=c=0, \hspace{5mm} b=-\frac{1}{2}, \hspace{5mm} d=\frac{1}{16}
\end{equation}
satisfy 5-brane solutions $c=0$ and $q=0$ in (\ref{condition cq in 2.2}). 
\par The global structure of the gauge group that forms in F-theory compactification on the extremal K3 surface (\ref{fibration [212] in 3.2.1}) was discussed in \cite{KRES}. It is
\begin{equation}
E_8^2\times SU(3).
\end{equation}
\par An attractive K3 surface $S_{[2 \hspace{1mm} 0 \hspace{1mm} 2]}$ admits an extremal elliptic fibration with the singularity type $E_8^2A_1^2$. The Weierstrass equation of this fibration is \cite{Shioda2008}
\begin{equation}
\label{fibration another [202] in 3.2.1}
y^2=x^3-3t^4\, x+t^5+t^7.
\end{equation}
In the homogeneous form, this is given by
\begin{equation}
\label{fibration [202] in 3.2.1}
y^2=x^3-3t^4s^4\, x+t^5s^7+t^7s^5.
\end{equation}
The discriminant is (in homogeneous form)
\begin{equation}
\label{discriminant [202] in 3.2.1}
\Delta \sim t^{10}s^{10}\, (t-s)^2(t+s)^2.
\end{equation}
It can be seen from (\ref{fibration [202] in 3.2.1}) and (\ref{discriminant [202] in 3.2.1}) that two type $II^*$ fibers are at $[t:s]=[0:1]$ and $[t:s]=[1:0]$, and two type $I_2$ fibers are at $[t:s]=[-1:1]$ and $[t:s]=[1:1]$. 
\par We find that
\begin{equation}
a=-3, \hspace{5mm} b=c=0, \hspace{5mm} d=1
\end{equation}
satisfy 5-brane solutions $c=0$ and $q=0$ in (\ref{condition cq in 2.2}).
\par The global structure of the gauge group that forms in F-theory compactification on the extremal K3 surface (\ref{fibration [202] in 3.2.1}) can be found in \cite{KRES}. It is 
\begin{equation}
E_8^2\times SU(2)^2.
\end{equation}

\par Now, we state the main result of this Section \ref{sssec3.2.1}. We show that the four extremal K3 surfaces in the moduli of K3 surfaces with the $E_8E_7$ singularity that we discussed previously can be viewed as special deformations of the stable degenerations in which singular fibers collide and 7-branes coincide. This shows that the non-geometric heterotic strings include the effect of coincident 7-branes and the enhancement of non-Abelian gauge groups on the F-theory side.
\par We saw that the extremal K3 surface $S_{[6 \hspace{1mm} 0 \hspace{1mm} 2]}$ has the singular fibers of type $II^*, III^*, I_3, I_2$. It was proved in \cite{KRES} that for any pair of extremal rational elliptic surfaces with distinct complex structures (except for $X_{[II, \hspace{1mm} II^*]}$, $X_{[III, \hspace{1mm} III^*]}$, and $X_{[IV, \hspace{1mm} IV^*]}$), there exists some K3 surface that stably degenerates into that pair. (For {\it any} pair of rational elliptic surfaces with an {\it identical} complex structure, there exists a K3 surface that stably degenerates into that pair \cite{KRES}.) There exists an elliptic K3 surface $S_1$ that splits into $X_{[II^*, \hspace{1mm} 1, 1]}\amalg X_{[III^*, \hspace{1mm} 2, 1]}$ under the stable degeneration. Because the singular fibers of the resulting K3 surface are given by the sum of the pair of rational elliptic surfaces \cite{KRES}, the elliptic K3 surface $S_1$ admits singular fibers of types $II^*, III^* I_2, I_1, I_1, I_1$. We consider the limit of this K3 surface $S_1$ at which the type $I_2$ fiber and a type $I_1$ fiber collide and two type $I_1$ fibers collide. In this limit, these fibers are enhanced to a type $I_3$ fiber and type $I_2$ fiber, respectively. Because the K3 surface obtained in the limit has the singularity type $E_8E_7A_2A_1$, the classification result in \cite{SZ} implies that the K3 surface obtained by taking this limit is the extremal K3 surface $S_{[6 \hspace{1mm} 0 \hspace{1mm} 2]}$. From this, we deduce that the extremal K3 surface $S_{[6 \hspace{1mm} 0 \hspace{1mm} 2]}$ is obtained when 7-branes coincide through the deformation of the K3 surface $S_1$ yielded as the reverse of the stable degeneration of the sum $X_{[II^*, \hspace{1mm} 1, 1]}\amalg X_{[III^*, \hspace{1mm} 2, 1]}$. 
\par We can also consider another deformation of the elliptic K3 surface $S_1$. This time, let us consider the limit at which the type $I_2$ fiber and two type $I_1$ fibers collide. These are enhanced to a type $I_4$ fiber in this situation, and the resulting elliptic K3 surface obtained in this limit has singular fibers of types $II^*, III^*, I_4,$ and $I_1$. The corresponding singularity type is $E_8E_7A_3$, and therefore from the classification result in \cite{SZ} we conclude that the resulting surface is the extremal K3 surface $S_{[4 \hspace{1mm} 0 \hspace{1mm} 2]}$. This shows that the extremal K3 surface $S_{[4 \hspace{1mm} 0 \hspace{1mm} 2]}$ can also be obtained when 7-branes coincide as a deformation of the K3 surface $S_1$. 
\par The reverse of the stable degeneration of the sum of two extremal rational elliptic surfaces $X_{[II, \hspace{1mm} II^*]}\amalg X_{[II, \hspace{1mm} II^*]}$ yields an elliptic K3 surface $S_2$, the singular fibers of which are two type $II^*$ fibers and two type $II$ fibers. It was deduced in \cite{KRES} that the extremal K3 surface $S_{[2 \hspace{1mm} 1 \hspace{1mm} 2]}$ with the $E_8^2A_2$ singularity is obtained as a deformation of $S_2$, in the limit at which two type $II$ fibers collide. These are enhanced to a type $IV$ fiber. This limit corresponds to the situation in which 7-branes over which type $II$ fibers lie become coincident. 
\par The reverse of the stable degeneration of the sum of two extremal rational elliptic surfaces $X_{[II^*, \hspace{1mm} 1, 1]} \amalg X_{[II^*, \hspace{1mm} 1, 1]}$ yields an elliptic K3 surface $S_3$, the singular fibers of which are two type $II^*$ fibers and four type $I_1$ fibers. As deduced in \cite{KRES}, the extremal K3 surface $S_{[2 \hspace{1mm} 0 \hspace{1mm} 2]}$ with the $E_8^2A_1^2$ singularity is obtained as a deformation of $S_3$ in the limit at which two pairs of type $I_1$ fibers collide, and these are enhanced to two type $I_2$ fibers. 
\par This argument shows that the four extremal K3 surfaces are deformations of some elliptic K3 surface that represent the stable degenerations of the sums of 1/2 K3s, and these deformations are the effect of coincident 7-branes. 

\subsubsection{Extremal K3 with the $E_6^3$ singularity}
\label{sssec3.2.2}
\par An attractive K3 surface $S_{[2 \hspace{1mm} 1 \hspace{1mm} 2]}$ has an extremal elliptic fibration with the singularity type $E_6^3$ (elliptic fibration no. 5 in Table \ref{tab[212]fibration} in Appendix \ref{secA}). This is given by taking the Jacobian fibration of the Fermat-type hypersurface in $\P^2\times\P^1$ studied in \cite{K}. 
\par The Fermat-type hypersurface in $\P^2\times\P^1$ \cite{K} is a genus-one fibered K3 surface defined by the equation of bidegree (3,2)
\begin{equation}
\label{Fermat type general in 3.2.2}
(t-\alpha_1)(t-\alpha_2)\, X^3+(t-\alpha_3)(t-\alpha_4)\, Y^3+(t-\alpha_5)(t-\alpha_6)\, X^3,
\end{equation}
where $t$ denotes the inhomogeneous coordinate on $\P^1$ in $\P^2\times\P^1$, $[X:Y:Z]$ denotes the homogeneous coordinates on $\P^2$ in $\P^2\times\P^1$, and $\alpha_i$, $i=1, \cdots, 6$ are parameters. Here, we particularly consider the most enhanced situation
\begin{equation}
\alpha_1=\alpha_2, \hspace{5mm} \alpha_3=\alpha_4, \hspace{5mm} \alpha_5=\alpha_6.
\end{equation}
In this situation, the genus-one fibration (\ref{Fermat type general in 3.2.2}) becomes 
\begin{equation}
\label{Fermat type enhanced in 3.2.2}
(t-\alpha_1)^2\, X^3+(t-\alpha_3)^2\, Y^3+(t-\alpha_5)^2\, X^3.
\end{equation}
The Fermat-type genus-one fibration (\ref{Fermat type general in 3.2.2}) does not have a global section \cite{K}, but its Jacobian fibration admits a global section. A genus-one fibration and its Jacobian fibration have the same types of the singular fibers, and their discriminant loci are identical. The Jacobian fibration of the K3 genus-one fibration (\ref{Fermat type enhanced in 3.2.2}) is \cite{K}
\begin{equation}
\label{Jacobian enhanced in 3.2.2}
y^2=x^3- 3^3 \cdot 2^4\cdot (t-\alpha_1)^4(t-\alpha_3)^4(t-\alpha_5)^4.
\end{equation}
The singular fibers of the K3 genus-one fibration (\ref{Fermat type enhanced in 3.2.2}) and the Jacobian fibration (\ref{Jacobian enhanced in 3.2.2}) are three type $IV^*$ fibers \cite{K}. Thus, the Jacobian fibration (\ref{Jacobian enhanced in 3.2.2}) yields an extremal elliptic fibration of the attractive K3 surface $S_{[2 \hspace{1mm} 1 \hspace{1mm} 2]}$ with the singularity type $E_6^3$.
\par Being an attractive K3 surface, $S_{[2 \hspace{1mm} 1 \hspace{1mm} 2]}$ admits a Shioda--Inose structure, and thus it should admit another elliptic fibration, which includes the singularity type $E_8^2$. In fact, the elliptic fibration (\ref{fibration another [212] in 3.2.1}) that we discussed previously admits such a fibration. The Jacobian fibration (\ref{Jacobian enhanced in 3.2.2}) and the elliptic fibration (\ref{fibration another [212] in 3.2.1}) are two elliptic fibrations of the same attractive K3 surface $S_{[2 \hspace{1mm} 1 \hspace{1mm} 2]}$. Therefore, there exists a birational map that transforms the Jacobian fibration (\ref{Jacobian enhanced in 3.2.2}) into the elliptic fibration (\ref{fibration another [212] in 3.2.1}). Because the Jacobian fibration (\ref{Jacobian enhanced in 3.2.2}) and the elliptic fibration (\ref{fibration another [212] in 3.2.1}) are birational, the F-theory compactification on the Jacobian (\ref{Jacobian enhanced in 3.2.2}) relates to a non-geometric $E_8\times E_8$ heterotic string.
\par The reverse of the stable degeneration of the sum $X_{[IV, \hspace{1mm} IV^*]}\amalg X_{[IV, \hspace{1mm} IV^*]}$ of two identical extremal rational elliptic surfaces $X_{[IV, \hspace{1mm} IV^*]}$ yields an elliptic K3 surface, which we denote as $S_4$. The singular fibers of this surface consist of two type $IV^*$ fibers and two type $IV$ fibers \cite{KRES}. A deformation of $S_4$ obtained by taking the limit at which two type $IV$ fibers collide yields an elliptic K3 surface $S_{[2 \hspace{1mm} 1 \hspace{1mm} 2]}$ with the extremal fibration (\ref{Jacobian enhanced in 3.2.2}) \cite{KRES}. Thus, the Jacobian fibration (\ref{Jacobian enhanced in 3.2.2}) is obtained from the K3 surface $S_4$, given as the stable degeneration of the sum of two 1/2 K3s, as the effect of coincident 7-branes.
\par The Mordell--Weil group of the Jacobian fibration (\ref{Jacobian enhanced in 3.2.2}) is isomorphic to $\Z_3$ \cite{Nish, SZ}, and the gauge group that arises for this fibration is \cite{K}
\begin{equation}
E_6^3 / \Z_3.
\end{equation}
In Section \ref{sec4}, we consider the fibering of the Jacobian fibrations of the genus-one fibrations (\ref{Fermat type general in 3.2.2}) over $\P^1$, yielding an elliptic Calabi--Yau 3-fold.

\subsection{Applications to $SO(32)$ heterotic string}
\label{ssec3.3}
A birational map from K3 elliptic fibration the singularity type of which includes $E_8E_7$ to K3 surface with another elliptic fibration with a type $I_{10}^*$ fiber was discussed in \cite{MM}. Using the birational transformation
\begin{equation}
\label{birational map in 3.3}
x \rightarrow x^2t, \hspace{5mm} y \rightarrow x^2y, \hspace{5mm} t \rightarrow x,
\end{equation}
the K3 elliptic fibration (\ref{Weierstrass E8E7 in 2.2}) was transformed to K3 elliptic fibration with a type $I^*_{10}$ fiber (or worse) in \cite{MM}, given by
\begin{equation}
\label{Weierstrass SO(28) in 3.3}
y^2=x^3+(t^3+a\, t+b)\, x^2+ (ct+d)\, x.
\end{equation}
This fibration relates to the $SO(32)$ heterotic string. The discriminant of the Weierstrass equation (\ref{Weierstrass SO(28) in 3.3}) is, in the homogeneous form, \cite{MM}
\begin{equation}
\Delta \sim s^{16}\, (ct+ds)^2 \, \Big(t^6+2a\, t^4s^2+2b\, t^3s^3+a^2\, t^2s^4+(2ab-4c)\, ts^5+(b^2-4d)\, s^6 \Big).
\end{equation}
$[t:s]$ denotes the coordinate of the base $\P^1$.
\par We discuss the implications of the birational transformation (\ref{birational map in 3.3}) and the resulting Weierstrass form (\ref{Weierstrass SO(28) in 3.3}) with a type $I^*_{10}$ fiber (or worse) for extremal K3 surfaces studied in section \ref{ssec3.2}. 
\par When the birational transformation is applied to the extremal fibration (\ref{fibration [602] in 3.2.1}) of the K3 surface $S_{[6 \hspace{1mm} 0 \hspace{1mm} 2]}$, by plugging the values for $a,b,c,d$ (\ref{abcd for [602] in 3.2.1}) for the Weierstrass form (\ref{Weierstrass SO(28) in 3.3}), we obtain the following equation:
\begin{equation}
\label{so(28) for [602] in 3.3}
y^2=x^3+(t^3-\frac{15}{2\sqrt[3]{2}}\, t-\frac{5}{2})\, x^2+(3\, \sqrt[3]{4}\, t+10)x.
\end{equation}
The discriminant of the resulting Weierstrass form is 
\begin{equation}
\label{discriminant [602] in 3.3}
\Delta \sim s^{16}\, (3\sqrt[3]{4}\, t+10s)^2\, (t-\sqrt[3]{2}s)^2\, (2t+3\sqrt[3]{2}s)^3\, (2t-5\sqrt[3]{2}s).
\end{equation}
From the equations (\ref{so(28) for [602] in 3.3}) and (\ref{discriminant [602] in 3.3}), we deduce that the transformed fibration (\ref{so(28) for [602] in 3.3}) has the singular fibers of types $I_{10}^*, I_3, I_2, I_2, I_1$. Thus, the singularity type of the fibration (\ref{so(28) for [602] in 3.3}) is $D_{14}A_2A_1^2$. The rank of the singularity type is 18, therefore this fibration is also extremal. Birational map (\ref{birational map in 3.3}) does not alter the complex structure of a K3 surface, namely, the birational map (\ref{birational map in 3.3}) sends the attractive K3 surface $S_{[6 \hspace{1mm} 0 \hspace{1mm} 2]}$ to itself, but the fibration structure is not preserved under this map. 
\par Therefore, the obtained results predict that the attractive K3 surface $S_{[6 \hspace{1mm} 0 \hspace{1mm} 2]}$ has the extremal elliptic fibration (\ref{so(28) for [602] in 3.3}) with the singularity type $ D_{14}A_2A_1^2$. This agrees with the classification of the singularity types of the extremal K3 surfaces in \cite{SZ} (See No.211 in Table 2). 
\par The extremal K3 surface with the singularity type $ D_{14}A_2A_1^2$ has the Mordell--Weil group isomorphic to $\Z_2$ \cite{SZ}, therefore, the Mordell--Weil group of the elliptic fibration (\ref{so(28) for [602] in 3.3}) is isomorphic to $\Z_2$. Thus, the gauge group that forms for this fibration is 
\begin{equation}
SO(28) \times SU(3) \times SU(2)^2 / \Z_2.
\end{equation}
The non-Abelian part of the gauge group has rank 18. This suggests that perturbative description of the gauge group is not possible for the $SO(32)$ heterotic string (same as for heterotic string with unbroken $\mathfrak{e}_8\mathfrak{e}_7$ gauge algebra).

\par For an attractive K3 surface $S_{[2 \hspace{1mm} 1 \hspace{1mm} 2]}$ with the extremal fibration (\ref{fibration [212] in 3.2.1}) with the singularity type $E_8^2A_2$, the birational transformation (\ref{birational map in 3.3}) maps the fibration (\ref{fibration [212] in 3.2.1}) to the following fibration:
\begin{equation}
\label{so(28) [212] in 3.3}
y^2=x^3+(t^3-\frac{1}{2})\, x^2+ \frac{1}{16}\, x.
\end{equation}
The discriminant of the resulting fibration is 
\begin{equation}
\label{discriminant [212] in 3.3}
\Delta \sim s^{18}\, t^3\, (t^3-s^3).
\end{equation}
From the equations (\ref{so(28) [212] in 3.3}) and (\ref{discriminant [212] in 3.3}), we find that the resulting fibration (\ref{so(28) [212] in 3.3}) has the singular fibers of types $I_{12}^*, I_3, I_1, I_1, I_1$, and the singularity type of the fibration (\ref{so(28) [212] in 3.3}) is $D_{16}A_2$. The fibration (\ref{so(28) [212] in 3.3}) is also extremal. 
\par This shows that the attractive K3 surface $S_{[2 \hspace{1mm} 1 \hspace{1mm} 2]}$ has an elliptic fibration with the singularity type $D_{16}A_2$. This agrees with the results in \cite{Nish, SZ}. (See fibration no.2 in Table \ref{tab[212]fibration} in Appendix \ref{secA}, and No.216 in Table 2 in \cite{SZ}.) See also, e.g., \cite{Ut} for studies of the fibration (\ref{so(28) [212] in 3.3}) of $S_{[2 \hspace{1mm} 1 \hspace{1mm} 2]}$.
\par The Mordell--Weil group of the fibration (\ref{so(28) [212] in 3.3}) is isomorphic to $\Z_2$ \cite{Nish, SZ}, and the gauge group that forms for this fibration is 
\begin{equation}
SO(32) \times SU(3) / \Z_2.
\end{equation}
The rank of the non-Abelian part of the gauge group is 18, and the perturbative description of the gauge group for the $SO(32)$ heterotic string is not possible either for this case.

The birational transformation (\ref{birational map in 3.3}) applied to an attractive K3 surface $S_{[2 \hspace{1mm} 0 \hspace{1mm} 2]}$ with the extremal fibration (\ref{fibration another [202] in 3.2.1}) with the singularity type $E_8^2A_1^2$ yields the following fibration:
\begin{equation}
\label{so(28) [202] in 3.3}
y^2=x^3+(t^3-3\, t)\, x^2+ x.
\end{equation}
The discriminant of this fibration is 
\begin{equation}
\label{discriminant [202] in 3.3}
\Delta \sim s^{18}\, (t^2-4s^2)\, (t-s)^2(t+s)^2.
\end{equation}
From the equations (\ref{so(28) [202] in 3.3}) and (\ref{discriminant [202] in 3.3}), we find that the fibration (\ref{so(28) [202] in 3.3}) has the singular fibers of types $I_{12}^*, I_2, I_2, I_1, I_1$, and the singularity type of the fibration (\ref{so(28) [202] in 3.3}) is $D_{16}A_1^2$. Therefore, the fibration (\ref{so(28) [202] in 3.3}) is also extremal. 
\par These computations show that the attractive K3 surface $S_{[2 \hspace{1mm} 0 \hspace{1mm} 2]}$ has an elliptic fibration with the singularity type $D_{16}A_1^2$. This agrees with the results in \cite{Nish, SZ}. (See fibration no.3 in Table \ref{tab[202]fibration} in Appendix \ref{secA}, and No.215 in Table 2 in \cite{SZ}.) 
\par The Mordell--Weil group of the fibration (\ref{so(28) [202] in 3.3}) is isomorphic to $\Z_2$ \cite{Nish, SZ}, and the gauge group that forms is 
\begin{equation}
SO(32) \times SU(2)^2 / \Z_2.
\end{equation}
The rank of the non-Abelian part of the gauge group is 18, and the perturbative description of the gauge group for the $SO(32)$ heterotic string is not possible either for this case.

\section{Jacobian of Fermat-type Calabi-Yau 3-folds and F-theory}
\label{sec4}
In this Section, we consider the Jacobian fibrations of (3,2,2) hypersurfaces in $\P^2\times\P^1\times\P^1$, in order to construct a family of elliptically fibered Calabi--Yau 3-folds \footnote{Elliptic fibrations of 3-folds are discussed in \cite{Nak, DG, G}.}. This family admits a K3 fibration. Choosing a specific form of equations for elliptic Calabi--Yau 3-folds, this K3 fiber coincides with the Jacobian of the Fermat-type K3 hypersurface, as discussed in Section \ref{sssec3.2.2}. We discuss F-theory compactifications on the constructed elliptic Calabi--Yau 3-folds. 

\subsection{Fermat-type (3,2,2) hypersurface, the Jacobian fibration, and the discriminant locus}
\label{ssec4.1}
A (3,2,2) hypersurface in $\P^2\times\P^1\times\P^1$ has the trivial canonical bundle $K=0$, and therefore such a hypersurface yields a Calabi--Yau 3-fold. The projection onto $\P^1\times\P^1$ yields a genus-one fibration, and the projection onto $\P^1$ gives a K3 fibration. 
\par In particular, we consider the (3,2,2) hypersurfaces given by the following form of the equations:
\begin{equation}
\label{(3,2,2)-1 in 4.1}
(t-\alpha_1)(t-\alpha_2)\, f_1\, X^3+(t-\alpha_3)(t-\alpha_4)\, f_2\, Y^3+(t-\alpha_5)(t-\alpha_6)\, f_3\, Z^3=0,
\end{equation}
where $[X:Y:Z]$ denotes the homogeneous coordinates on $\P^2$ in $\P^2\times\P^1\times\P^1$, $t$ is the inhomogeneous coordinate on the first $\P^1$ in $\P^2\times\P^1\times\P^1$, and $f_1, f_2, f_3$ are degree 2 polynomials on the second $\P^1$ in $\P^2\times\P^1\times\P^1$. Furthermore, $\alpha_i$, $i=1, \cdots, 6$ denote points in $\P^1$. We call the (3,2,2) hypersurfaces of this form {\it Fermat-type} hypersurfaces. (Similar naming conventions were employed for the K3 hypersurfaces and Calabi--Yau 4-folds constructed in \cite{K, KCY4}.) K3 fibers of these Calabi--Yau 3-folds are Fermat-type K3 (3,2) hypersurfaces \cite{K}, as discussed in Section \ref{sssec3.2.2}. In particular, K3 fibers are genus-one fibered, but they lack a section \cite{K}. Splitting the degree 2 polynomials $f_1, f_2, f_3$ into linear factors, the equation (\ref{(3,2,2)-1 in 4.1}) can be rewritten as
\begin{equation}
\label{(3,2,2)-2 in 4.1}
\begin{split}
(t-\alpha_1)(t-\alpha_2)(u-\beta_1)(u-\beta_2)\, X^3+(t-\alpha_3)(t-\alpha_4)(u-\beta_3)(u-\beta_4)\, Y^3 & \\
+(t-\alpha_5)(t-\alpha_6)(u-\beta_5)(u-\beta_6)\, Z^3 & =0.
\end{split}
\end{equation}
By $u$ we denote the inhomogeneous coordinate on the second $\P^1$ in $\P^2\times\P^1\times\P^1$, and $\beta_i$, $i=1, \cdots, 6$ denote points in this $\P^1$. 
\par Similar argument to that in \cite{KCY4} implies that the genus-one fibration (\ref{(3,2,2)-2 in 4.1}) does not admit a global section. If the genus-one fibration (\ref{(3,2,2)-2 in 4.1}) admits a section, this restricts to K3 fibers, giving a section to K3 fibers. K3 fibers are Fermat-type K3 (3,2) hypersurfaces (\ref{Fermat type general in 3.2.2}), which do not admit a section, as explained previously. Therefore, this implies a contradiction. By an argument similar to those in \cite{K, KCY4, Kdisc}, the genus-one fibration (\ref{(3,2,2)-2 in 4.1}) admits a 3-section \footnote{Therefore, a discrete $\Z_3$ gauge symmetry \cite{KMOPR, CDKPP, Kdisc} arises in F-theory compactifications on the genus-one fibered Calabi--Yau 3-folds (\ref{(3,2,2)-1 in 4.1}).}. 
\par By taking the Jacobian fibration of the genus-one fibrations (\ref{(3,2,2)-2 in 4.1}), we construct a family of elliptically fibered Calabi--Yau 3-folds with a section. The Jacobian fibration of the genus-one fibration (\ref{(3,2,2)-2 in 4.1}) is given by the following equation:
\begin{equation}
\label{J(3,2,2) in 4.1}
X^3+Y^3+\Pi_{i=1}^6(t-\alpha_i) \, \Pi_{j=1}^6(u-\beta_j)\, Z^3=0.
\end{equation}
This equation transforms into the Weierstrass form as
\begin{equation}
\label{Weierstrass J(3,2,2) in 4.1}
y^2=x^3- 3^3 \cdot 2^4 \cdot \Pi_{i=1}^6(t-\alpha_i)^2 \, \Pi_{j=1}^6(u-\beta_j)^2.
\end{equation}
The Jacobian fibrations (\ref{Weierstrass J(3,2,2) in 4.1}) yield elliptically fibered Calabi--Yau 3-folds. In Section \ref{sec4}, we discuss F-theory compactifications on these Calabi--Yau 3-folds. These compactifications yield a six-dimensional theory with $N=1$ supersymmetry.
\par The discriminant of the Weierstrass equation (\ref{Weierstrass J(3,2,2) in 4.1}) is given by 
\begin{equation}
\Delta \sim \Pi_{i=1}^6 (t-\alpha_i)^4 \, \Pi_{j=1}^6 (u-\beta_j)^4.
\end{equation}
Then, 7-branes are wrapped on the irreducible components of the discriminant locus $\Delta=0$. There are 12 components, which we denote by $A_i$, $i=1, \cdots, 6$, and $B_j$, $j=1, \cdots, 6$. These are given by the following equations:
\begin{eqnarray}
A_i & = & \{t=\alpha_i\} \hspace{5mm} (i=1, \cdots, 6) \\ \nonumber
B_j & = & \{u=\beta_j\} \hspace{5mm} (j=1, \cdots, 6).
\end{eqnarray}
Because $A_i$ is isomorphic to $\{{\rm pt}\} \times \P^1$ and $B_j$ is isomorphic to $\P^1\times \{{\rm pt}\}$, these are isomorphic to $\P^1$.

\subsection{Gauge groups}
\label{ssec4.2}
\subsubsection{Singular fibers and non-Abelian gauge groups}
\label{sssec4.2.1}
We determine the non-Abelian gauge groups that form on the 7-branes in F-theory compactifications on the Jacobian fibrations (\ref{Weierstrass J(3,2,2) in 4.1}). In a generic case in which $\alpha_i$ and $\beta_j$ are mutually distinct, all singular fibers are of type $IV$. When two $\alpha_i$'s become coincident, say $\alpha_1=\alpha_2$, the corresponding 7-branes $A_1$ and $A_2$ coincide, and the singular fiber on the 7-brane $A_1$ is enhanced to type $IV^*$. Similarly, when two $\beta_j$'s are coincident, say $\beta_1=\beta_2$, the 7-branes $B_1$ and $B_2$ coincide, and the singular fiber on the 7-brane $B_1$ is enhanced to type $IV^*$. (See Table \ref{tabvanishingorder in 2.1} in Section \ref{ssec2.1}.) When three of the $\alpha_i$'s become coincident, the Calabi--Yau condition is broken \cite{K, KCY4}. A similar statement holds for the $\beta_j$'s. 
\par From the equation (\ref{Weierstrass J(3,2,2) in 4.1}), we find that both the type $IV$ fiber and type $IV^*$ fiber are split \cite{BIKMSV}. Therefore, the non-Abelian gauge group that forms is $SU(3)$ for a type $IV$ fiber and $E_6$ for a type $IV^*$ fiber. The results are summarized in Table \ref{tab gauge group in 4.2.1}. 

\begingroup
\renewcommand{\arraystretch}{1.5}
\begin{table}[htb]
\begin{center}
  \begin{tabular}{|c|c|c|} \hline
Component & fiber type & non-Abel. Gauge Group \\ \hline
$A_{1,\cdots, 6}$ & $IV$ & $SU(3)$ \\ \hline
$A_{1,\cdots, 6}$ & $IV^*$ & $E_6$ \\ \hline 
$B_{1,\cdots, 6}$ & $IV$ & $SU(3)$ \\ \hline  
$B_{1,\cdots, 6}$ & $IV^*$ & $E_6$ \\ \hline
\end{tabular}
\caption{Singular fiber types and the gauge groups on the discriminant components.} 
\label{tab gauge group in 4.2.1}
\end{center}
\end{table}  
\endgroup

\par Generically, there are 12 discriminant components with type $IV$ fibers, and thus the non-Abelian gauge group that forms is 
\begin{equation}
SU(3)^{12}.
\end{equation}
When three pairs of $\alpha$ coincide, 
\begin{equation}
\alpha_1=\alpha_2, \hspace{5mm} \alpha_3=\alpha_4, \hspace{5mm} \alpha_5=\alpha_6,
\end{equation}
the Jacobian fibration (\ref{J(3,2,2) in 4.1}) becomes 
\begin{equation}
X^3+Y^3+(t-\alpha_1)^2(t-\alpha_3)^2(t-\alpha_5)^2\, \Pi_{j=1}^6(u-\beta_j)\, Z^3=0.
\end{equation}
The Weierstrass form is then
\begin{equation}
\label{Weierstrass J(3,2,2) enhanced in 4.2.1}
y^2=x^3- 3^3\cdot 2^4\cdot (t-\alpha_1)^4(t-\alpha_3)^4(t-\alpha_5)^4 \, \Pi_{j=1}^6(u-\beta_j)^2.
\end{equation}
In this situation, the K3 fibers of the Jacobian (\ref{Weierstrass J(3,2,2) enhanced in 4.2.1}) constitute the extremal K3 surface (\ref{Jacobian enhanced in 3.2.2}), as discussed in Section \ref{sssec3.2.2}. The 7-branes coincide as $A_1=A_2$, $A_3=A_4$, and $A_5=A_6$, and the non-Abelian gauge group that forms in enhanced to 
\begin{equation}
\label{global gauge in 4.2.1}
E_6^3 \times SU(3)^{6}.
\end{equation}
We assumed that the $\beta$'s are mutually distinct. When pairs of $\beta$ are coincident, the corresponding 7-branes coincide. Then, $SU(3)^2$ is enhanced to $E_6$, and the gauge group (\ref{global gauge in 4.2.1}) is further enhanced if this occurs.

\subsubsection{Consistency check}
\label{sssec4.2.2}
By an argument similar to those given in \cite{K, KCY4}, smooth genus-one fibers of the genus-one fibration (\ref{(3,2,2)-2 in 4.1}) are the Fermat curves. Thus, they have j-invariant 0. This means that generic fibers over the base surface have j-invariant 0, and this forces the singular fibers to have j-invariant 0. This condition strongly constrains possible fiber types of the singular fibers. Because a genus-one fibration and its Jacobian fibration have the identical types and configurations of the singular fibers, the same condition is imposed on the singular fibers of the Jacobian (\ref{J(3,2,2) in 4.1}).
\par As in Table \ref{tabfibertypes in 2.1} in Section \ref{ssec2.1}, the singular fibers with j-invariant 0 are: type $II$, $II^*$, $IV$, $IV^*$, and type $I_0^*$ (type $I_0^*$ fiber is also included, because type $I_0^*$ fiber can have j-invariant 0.) We deduced in Section \ref{sssec4.2.1} that the singular fibers have types $IV$ and $IV^*$. We confirm that this result agrees with the constraint imposed. Monodromies of order 3 characterize the gauge groups.

\subsection{Models without U(1) and their Mordell-Weil groups}
\label{ssec4.3}
Here, we discuss the particular Jacobian fibration (\ref{Weierstrass J(3,2,2) enhanced in 4.2.1}). The K3 fiber of this Jacobian fibration is the extremal K3 surface $S_{[2 \hspace{1mm} 1 \hspace{1mm} 2]}$ with the singularity type $E_6^3$ (\ref{Jacobian enhanced in 3.2.2}), and therefore the Mordell--Weil group of the K3 fiber is isomorphic to $\Z_3$ \cite{Nish, SZ}. By an argument similar to those given in \cite{KCY4, Kdisc}, we can consider the specialization to the K3 fiber to find that the Mordell--Weil group of the Jacobian fibration (\ref{Weierstrass J(3,2,2) enhanced in 4.2.1}) is isomorphic to the Mordell--Weil group of the K3 fiber. Thus, the Mordell--Weil group of the Jacobian fibration (\ref{Weierstrass J(3,2,2) enhanced in 4.2.1}) is isomorphic to $\Z_3$:
\begin{equation}
MW \cong \Z_3.
\end{equation}
The global structure \cite{AspinwallGross, AMrational, MMTW} of the gauge group that forms on the 7-branes is
\begin{equation}
\label{gauge group in 4.3}
E_6^3 \times SU(3)^6 / \Z_3.
\end{equation}
The F-theory compactifications on the Jacobians (\ref{Weierstrass J(3,2,2) enhanced in 4.2.1}) do not have a $U(1)$ gauge group.

\subsection{Matter spectra}
In this section, we deduce candidates for the matter spectra on F-theory compactifications on the Jacobian Calabi--Yau 3-folds. F-theory compactifications on the Jacobian Calabi--Yau 3-folds yield six-dimensional theories with $N=1$ supersymmetry. We confirm that the deduced candidates for matter satisfy the anomaly cancellation condition \cite{GSW6d, Sagnotti, Erler, Sch6d}. Because the base of the Jacobian Calabi--Yau 3-folds (\ref{Weierstrass J(3,2,2) in 4.1}) is isomorphic to $\P^1\times\P^1$, 
\begin{equation}
T=h^{1,1}(\P^1\times\P^1)-1=2-1=1.
\end{equation}
Thus, $H$ and $V$ are required to satisfy the following equation:
\begin{equation}
\label{H-V in 4.4}
H-V=273-29=244.
\end{equation}
\par In particular, We consider the specific situations in which three pairs of the $\alpha$'s coincide:
\begin{equation}
\label{three alpha in 4.4}
\alpha_1=\alpha_2, \hspace{5mm} \alpha_3=\alpha_4, \hspace{5mm} \alpha_5=\alpha_6,
\end{equation}
as described previously in Section \ref{sssec4.2.1}. As we saw in Section \ref{ssec4.3}, in this case the Mordell--Weil ranks of the Jacobian fibrations (\ref{Weierstrass J(3,2,2) enhanced in 4.2.1}) are 0, and the Mordell--Weil groups are isomorphic to $\Z_3$. 
\par First, we consider a generic case in which the $\beta$'s are mutually distinct. The discriminant of the Jacobian (\ref{Weierstrass J(3,2,2) enhanced in 4.2.1}) is given by
\begin{equation}
\Delta \sim (t-\alpha_1)^8 \, (t-\alpha_3)^8 \, (t-\alpha_5)^8 \, \Pi_{j=1}^6 (u-\beta_j)^4.
\end{equation}
and the 7-branes intersect at 18 points:
\begin{equation}
(\alpha_i, \beta_j) \hspace{5mm} (i=1, 3, 5, \hspace{1mm} j=1, \cdots, 6).
\end{equation}
For all 18 intersections, the 7-branes over which type $IV^*$ fibers lie intersect those over which type $IV$ fibers lie. The corresponding singularity types are $E_6$ and $A_2$.
\par We saw in Section \ref{ssec4.3} that the gauge group that arises is $E_6^3 \times SU(3)^6 / \Z_3$. Therefore,
\begin{equation}
V=78\times 3+8\times 6=282.
\end{equation}
Thus, from the anomaly cancellation condition (\ref{H-V in 4.4}) we must have 
\begin{equation}
H=282+244=526.
\end{equation}
\par Although there appear to be three parameters of $\alpha$ in the equation (\ref{Weierstrass J(3,2,2) enhanced in 4.2.1}), $\alpha_1, \alpha_3, \alpha_5$, using automorphism of $\P^1$ we may assume that they are $0, 1, \infty$. Thus, these three parameters are superfluous, and they do not count as the parameters for the deformation of the complex structure. By a similar reasoning, we may also assume that $\beta_1, \beta_2, \beta_3$ are equal to $0, 1, \infty$, respectively, using automorphism of $\P^1$. Therefore, the actual parameters of the deformation of the complex structure of the Jacobian Calabi--Yau 3-fold (\ref{Weierstrass J(3,2,2) enhanced in 4.2.1}) are $\beta_4, \beta_5, \beta_6$, namely, the number of the parameters of the deformation of the complex structure is three. Thus, the number of the neutral hypermultiplets that arise from the deformation of the complex structure is 
\begin{equation}
H^0=1+3=4.
\end{equation}
We used $H^0$ to denote the number of the neutral hypermultiplets. It follows that the number of the hypermultiplets that arise from the 18 intersections of the 7-branes is
\begin{equation}
526-4=522.
\end{equation}
When the following matter arise at each of the 18 intersections
\begin{equation}
{\bf 27}\oplus {\bf 1}\oplus {\bf 1},
\end{equation}
the anomaly cancellation condition is satisfied:
\begin{equation}
(27+1+1)\times 18=522.
\end{equation}
Therefore, we deduce that 18 ${\bf 27}\oplus {\bf 1}\oplus {\bf 1}$ and 4 uncharged neutral hypermultiplets yield candidate matter spectrum in F-theory compactification on the Jacobian Calabi--Yau 3-fold (\ref{Weierstrass J(3,2,2) enhanced in 4.2.1}).

\par Second, we consider the case in which a pair of $\beta$ coincides:
\begin{equation}
\beta_1=\beta_2.
\end{equation}
Then, the Weierstrass equation of the Jacobian Calabi--Yau 3-fold is given by
\begin{equation}
\label{Weierstrass J(3,2,2) enhanced (4,4) in 4.4}
y^2=x^3- 3^3\cdot 2^4\cdot (t-\alpha_1)^4(t-\alpha_3)^4(t-\alpha_5)^4 \, (u-\beta_1)^4 \, \Pi_{j=3}^6(u-\beta_j)^2.
\end{equation}
The discriminant of the Jacobian (\ref{Weierstrass J(3,2,2) enhanced (4,4) in 4.4}) is given by
\begin{equation}
\Delta \sim (t-\alpha_1)^8 \, (t-\alpha_3)^8 \, (t-\alpha_5)^8 \, (u-\beta_1)^8 \, \Pi_{j=3}^6 (u-\beta_j)^4.
\end{equation}
and the 7-branes intersect at 15 points:
\begin{equation}
(\alpha_i, \beta_j) \hspace{5mm} (i=1, 3, 5, \hspace{1mm} j=1, 3, 4, 5, 6).
\end{equation}
For the 3 intersections $(\alpha_i, \beta_j)$, $i=1,3,5, \, j=1$, the 7-branes over which type $IV^*$ fibers lie intersect with those over which type $IV^*$ fibers lie. The corresponding singularity types are $E_6$ and $E_6$. The 7-branes over which type $IV^*$ fibers lie intersect those over which type $IV$ fibers lie for the remaining 12 intersections $(\alpha_i, \beta_j)$, $i=1,3,5, \, j=3,4,5,6$. The corresponding singularity types are $E_6$ and $A_2$.
\par The 7-branes $B_1$ and $B_2$ coincide, and the fiber type on $B_1$ is enhanced to type $IV^*$. Thus, the gauge group that arises in F-theory compactification on the Jacobian Calabi--Yau 3-fold (\ref{Weierstrass J(3,2,2) enhanced (4,4) in 4.4}) is 
\begin{equation}
E_6^4 \times SU(3)^4 / \Z_3.
\end{equation}
Therefore, we have that
\begin{equation}
V=78\times 4+8\times 4=344,
\end{equation}
and the anomaly cancellation condition (\ref{H-V in 4.4}) requires that 
\begin{equation}
H=344+244=588.
\end{equation}
\par By a reasoning similar to that stated previously, we may assume that $\beta_1, \beta_3, \beta_4$ are equal to $0, 1, \infty$, respectively, using automorphism of $\P^1$. Therefore, the parameters for the deformation of the complex structure are $\beta_5, \beta_6$, and the number of the neutral hypermultiplet that arises is
\begin{equation}
H^0=1+2=3.
\end{equation}
It follows that the number of the hypermultiplets that arise from the 15 intersections of the 7-branes is
\begin{equation}
588-3=585.
\end{equation}
When the following matter arise at each of the 3 intersections $(\alpha_i, \beta_j)$, $i=1,3,5, \, j=1$ 
\begin{equation}
{\bf 78}\oplus {\bf 1},
\end{equation}
and the following matter arise at each of the remaining 12 intersections $(\alpha_i, \beta_j)$, $i=1,3,5, \, j=3,4,5,6$
\begin{equation}
{\bf 27}\oplus {\bf 1}\oplus {\bf 1},
\end{equation}
we find that the anomaly cancellation condition is satisfied:
\begin{equation}
(78+1)\times 3+(27+1+1)\times 12=585.
\end{equation}
Thus, we confirm that 3 ${\bf 78}\oplus {\bf 1}$, 12 ${\bf 27}\oplus {\bf 1}\oplus {\bf 1}$ and 3 uncharged neutral hypermultiplets yield candidate matter spectrum in F-theory compactification on the Jacobian Calabi--Yau 3-fold (\ref{Weierstrass J(3,2,2) enhanced (4,4) in 4.4}).

\par Next, we discuss the case in which two pairs of $\beta$ coincide:
\begin{equation}
\beta_1=\beta_2, \hspace{5mm} \beta_3=\beta_4.
\end{equation}
In this case, the Weierstrass equation of the Jacobian Calabi--Yau 3-fold is given by the following equation
\begin{equation}
\label{Weierstrass J(3,2,2) enhanced (5,2) in 4.4}
y^2=x^3- 3^3\cdot 2^4\cdot (t-\alpha_1)^4(t-\alpha_3)^4(t-\alpha_5)^4 \, (u-\beta_1)^4(u-\beta_3)^4 \, (u-\beta_5)^2 (u-\beta_6)^2.
\end{equation}
The discriminant is
\begin{equation}
\Delta \sim (t-\alpha_1)^8 \, (t-\alpha_3)^8 \, (t-\alpha_5)^8 \, (u-\beta_1)^8 \, (u-\beta_3)^8 \, (u-\beta_5)^4 \, (u-\beta_6)^4.
\end{equation}
\par The 7-branes intersect at 12 points:
\begin{equation}
(\alpha_i, \beta_j) \hspace{5mm} (i=1, 3, 5, \hspace{1mm} j=1, 3, 5, 6).
\end{equation}
For the 6 intersections $(\alpha_i, \beta_j)$, $i=1,3,5, \, j=1, 3$, the 7-branes over which type $IV^*$ fibers lie intersect with those over which type $IV^*$ fibers lie. The corresponding singularity types are both $E_6$. The 7-branes over which type $IV^*$ fibers lie intersect those over which type $IV$ fibers lie for the remaining 6 intersections $(\alpha_i, \beta_j)$, $i=1,3,5, \, j=5,6$. The corresponding singularity types are $E_6$ and $A_2$.
\par The 7-branes $B_1$ and $B_2$ coincide, as do $B_3$ and $B_4$, and the singular fibers over these 7-branes are of type $IV^*$. Thus, the gauge group that arises in F-theory compactification on the Jacobian Calabi--Yau 3-fold (\ref{Weierstrass J(3,2,2) enhanced (5,2) in 4.4}) is 
\begin{equation}
E_6^5 \times SU(3)^2 / \Z_3.
\end{equation}
We have that
\begin{equation}
V=78\times 5+8\times 2=406,
\end{equation}
and the anomaly cancellation condition (\ref{H-V in 4.4}) requires that the number of the hypermultiplets is
\begin{equation}
H=406+244=650.
\end{equation}
\par By a reasoning similar to that stated previously, using automorphism of $\P^1$ we may assume that $\beta_1, \beta_3, \beta_5$ are equal to $0, 1, \infty$, respectively. Therefore, the parameter for the deformation of the complex structure is $\beta_6$, and the number of the neutral hypermultiplet that arises is
\begin{equation}
H^0=1+1=2.
\end{equation}
It follows that the number of the hypermultiplets that arise from the 12 intersections of the 7-branes is
\begin{equation}
650-2=648.
\end{equation}
When the following matter arise at each of the 6 intersections $(\alpha_i, \beta_j)$, $i=1,3,5, \, j=1, 3$ 
\begin{equation}
{\bf 78}\oplus {\bf 1},
\end{equation}
and the following matter arise at each of the remaining 6 intersections $(\alpha_i, \beta_j)$, $i=1,3,5, \, j=5,6$
\begin{equation}
{\bf 27}\oplus {\bf 1}\oplus {\bf 1},
\end{equation}
the anomaly cancellation condition is satisfied:
\begin{equation}
(78+1) \times 6+(27+1+1)\times 6=648.
\end{equation}
Therefore, we find that 6 ${\bf 78}\oplus {\bf 1}$, 6 ${\bf 27}\oplus {\bf 1}\oplus {\bf 1}$ and 2 uncharged neutral hypermultiplets yield candidate matter spectrum on F-theory compactification on the Jacobian Calabi--Yau 3-fold (\ref{Weierstrass J(3,2,2) enhanced (5,2) in 4.4}).

\par Finally, we discuss the case in which three pairs of the $\beta$'s coincide:
\begin{equation}
\beta_1=\beta_2, \hspace{5mm} \beta_3=\beta_4, \hspace{5mm} \beta_5=\beta_6.
\end{equation}
In this case, the Weierstrass form of the Jacobian Calabi--Yau 3-fold is given by
\begin{equation}
\label{Weierstrass J(3,2,2) enhanced (6,0) in 4.4}
y^2=x^3- 3^3\cdot 2^4\cdot (t-\alpha_1)^4(t-\alpha_3)^4(t-\alpha_5)^4 \, (u-\beta_1)^4(u-\beta_3)^4 \, (u-\beta_5)^4.
\end{equation}
The discriminant is
\begin{equation}
\Delta \sim (t-\alpha_1)^8 \, (t-\alpha_3)^8 \, (t-\alpha_5)^8 \, (u-\beta_1)^8 \, (u-\beta_3)^8 \, (u-\beta_5)^8.
\end{equation}
\par The 7-branes intersect in 9 points:
\begin{equation}
(\alpha_i, \beta_j) \hspace{5mm} (i=1, 3, 5, \hspace{1mm} j=1, 3, 5).
\end{equation}
For all 9 intersections $(\alpha_i, \beta_j)$, $i=1,3,5, \, j=1, 3, 5$, the 7-branes over which type $IV^*$ fibers lie intersect with those over which type $IV^*$ fibers lie. The corresponding singularity types are both $E_6$. 
\par The 7-branes $B_1$ and $B_2$, $B_3$ and $B_4$, $B_5$ and $B_6$ coincide, and the singular fibers over these 7-branes are of type $IV^*$. Thus, the gauge group that arises in F-theory compactification on the Jacobian Calabi--Yau 3-fold (\ref{Weierstrass J(3,2,2) enhanced (6,0) in 4.4}) is 
\begin{equation}
E_6^6/ \Z_3.
\end{equation}
We have 
\begin{equation}
V=78\times 6=468.
\end{equation}
The anomaly cancellation condition (\ref{H-V in 4.4}) requires that the number of the hypermultiplets is
\begin{equation}
H=468+244=712.
\end{equation}
\par By a reasoning similar to that stated previously, using automorphism of $\P^1$ we may assume that $\beta_1, \beta_3, \beta_5$ are equal to $0, 1, \infty$, respectively. Therefore, the complex structure of the Jacobian Calabi--Yau 3-fold for this case is fixed. Thus, the number of the neutral hypermultiplet that arises is
\begin{equation}
H^0=1+0=1.
\end{equation}
It follows that the number of the hypermultiplets that arise from the 12 intersections of the 7-branes is
\begin{equation}
712-1=711.
\end{equation}
When the matter  
\begin{equation}
{\bf 78}\oplus {\bf 1}
\end{equation}
arise at each of the 9 intersections $(\alpha_i, \beta_j)$, $i=1,3,5, \, j=1, 3, 5$
the anomaly cancellation condition is satisfied:
\begin{equation}
(78+1) \times 9=711.
\end{equation}
Therefore, we deduce that 9 ${\bf 78}\oplus {\bf 1}$ and 1 uncharged neutral hypermultiplets yield candidate matter spectrum in F-theory compactification on the Jacobian Calabi--Yau 3-fold (\ref{Weierstrass J(3,2,2) enhanced (6,0) in 4.4}).

\vspace{5mm}

\par We expect that each of the first and the last matter spectra (18 $\times$ (${\bf 27}\oplus{\bf 1}\oplus{\bf 1}$) and 9 $\times$ (${\bf 78}\oplus {\bf 1}$)) is the unique choice that is free of anomaly. Concerning the first matter spectrum, we saw that 7-branes intersect at 18 points. A symmetry argument suggests that the identical matter arise from these 18 points. Thus, the dimension of the matter representation at each of the 18 intersections of the 7-branes should be 29, for the anomaly to be cancelled. An $E_6$ singularity and an $A_2$ singularity collide at each of these intersections. Therefore, it is natural to expect that {\bf 27} of $E_6$ arises at the intersections. It follows that the matter representation ${\bf 27}\oplus{\bf 1}\oplus{\bf 1}$ at each of the 18 intersections of the 7-branes is the unique choice. 
\par As to the last matter spectrum, the 7-branes intersect at 9 points. A symmetry argument again suggests that the identical matter arise from these 9 points, and the dimension of the matter representation at each of these points should be 79. Then the representation ${\bf 78}\oplus {\bf 1}$ is the unique natural choice. (The remaining two choices, ${\bf 27}+52 \times {\bf 1}$, or $2 \times {\bf 27}+ 25\times {\bf 1}$, do not seem natural.)
\par The second and the third matter spectra, 3$\times$(${\bf 78}\oplus {\bf 1}$) + 12$\times$(${\bf 27}\oplus {\bf 1}\oplus {\bf 1}$), and 6$\times$(${\bf 78}\oplus {\bf 1}$) + 6$\times$(${\bf 27}\oplus {\bf 1}\oplus {\bf 1}$), are not the unique choices. For example, 3$\times$(${\bf 27}\oplus {\bf 27}\oplus {\bf 1}$) + 12$\times$(${\bf 27}\oplus {\bf 8}$) is also a potentially viable matter spectrum that is free of anomaly in F-theory compactification on the Jacobian Calabi--Yau 3-fold (\ref{Weierstrass J(3,2,2) enhanced (4,4) in 4.4}). 6$\times${\bf 27} + 6$\times$({\bf 27}, {\bf 3}), and 6$\times${\bf 78} + 6$\times$(${\bf 27}\oplus {\bf 3}$) can also be viable matter spectra that are free of anomaly in F-theory compactification on the Jacobian Calabi--Yau 3-fold (\ref{Weierstrass J(3,2,2) enhanced (5,2) in 4.4}).
\par If we assume that matter representation ${\bf 27}\oplus {\bf 1}\oplus {\bf 1}$ arises from the collision of $E_6$ and $A_2$ singularities of the Jacobian Calabi--Yau 3-folds, and ${\bf 78}\oplus {\bf 1}$ arises from the collision of two $E_6$ singularities \footnote{The collision of two $E_6$ singularities in the context of four-dimensional conformal matter in F-theory was discussed in \cite{AHMT1803}. See also, e.g., \cite{AM, MT1201} for discussion of the collision of two $E_6$ singularities in F-theory.}, the anomaly cancels in F-theory compactifications on the four families of the Jacobian Calabi--Yau 3-folds (\ref{Weierstrass J(3,2,2) enhanced in 4.2.1}), (\ref{Weierstrass J(3,2,2) enhanced (4,4) in 4.4}), (\ref{Weierstrass J(3,2,2) enhanced (5,2) in 4.4}), (\ref{Weierstrass J(3,2,2) enhanced (6,0) in 4.4}). Determining whether these interpretations indeed give the matter spectra by resolving the singularities of the Jacobian Calabi--Yau 3-folds constructed in this note can be a direction of future studies.

\section{Conclusions}
\label{sec5}
We analyzed some points in the eight-dimensional F-theory moduli at which the ranks of the gauge groups are enhanced to 18, and their non-geometric heterotic duals. On the F-theory side, these models are understood as deformations of the stable degenerations, and non-Abelian gauge groups are enhanced as the effect of coincident 7-branes. This suggests that non-geometric heterotic strings include the effect of coincident 7-branes. In the heterotic language, this corresponds to the effect of the insertion of 5-branes.
\par We particularly considered $E_8\times E_8$ heterotic strings, but by applying a birational transformation from a K3 surface with type $II^*$ and $III^*$ fibers to a K3 surface with a type $I^*_{10}$ fiber \cite{ClingherDoran2012, MMS, MM}, we also discussed $SO(32)$ heterotic string. Considering F-theory dual of $SO(32)$ heterotic string, we deduced the Weierstrass equation of an extremal K3 surface.
\par Non-geometric heterotic duals of F-theory on K3 surfaces with the Picard number 19 with the singularity rank 17 also do not allow for perturbative interpretations of the gauge groups that arise on the heterotic side. This case is not studied in this note. It can be interesting to investigate F-theory on such K3 surfaces and the non-geometric heterotic duals.
\par In Section \ref{sec4}, by taking the Jacobian fibration of the Fermat-type (3,2,2) hypersurfaces we constructed elliptically fibered Calabi--Yau 3-folds, and we studied F-theory compactifications on these Calabi--Yau 3-folds. Highly enhanced gauge groups arise in these compactifications. We deduced candidate matter spectra directly from the global defining equation of the Calabi--Yau 3-folds. These Calabi--Yau 3-folds admit a K3 fibration. They include special cases in which K3 fibrations become extremal. In this situation, F-theory compactifications on the K3 fibers have non-geometric heterotic duals (by applying a birational map if necessary). Determining whether six-dimensional F-theory on Calabi--Yau 3-folds have heterotic duals is a likely direction of future studies.

\section*{Acknowledgments}

We would like to thank Shun'ya Mizoguchi for discussions. This work is partially supported by Grant-in-Aid for Scientific Research {\#}16K05337 from the Ministry of Education, Culture, Sports, Science and Technology of Japan.

\newpage
\appendix
\section{Elliptic fibrations of attractive K3 $S_{[2 \hspace{1mm} 1 \hspace{1mm} 2]}$ and $S_{[2 \hspace{1mm} 0 \hspace{1mm} 2]}$}
\label{secA}
The types of the elliptic fibrations of attractive K3 surfaces with the discriminant three and four, $S_{[2 \hspace{1mm} 1 \hspace{1mm} 2]}$ and $S_{[2 \hspace{1mm} 0 \hspace{1mm} 2]}$, were classified, and the Mordell--Weil groups of the fibrations were determined in \cite{Nish}. We list the types of the elliptic fibrations and the Mordell--Weil groups of $S_{[2 \hspace{1mm} 1 \hspace{1mm} 2]}$ and $S_{[2 \hspace{1mm} 0 \hspace{1mm} 2]}$ deduced in \cite{Nish} in Table \ref{tab[212]fibration} and Table \ref{tab[202]fibration}.

\begingroup
\renewcommand{\arraystretch}{1.1}
\begin{table}[htb]
\centering
  \begin{tabular}{|c|c|c|} \hline
$
\begin{array}{c}
\mbox{Elliptic fibrations of}\\
\mbox{$S_{[2 \hspace{1mm} 1 \hspace{1mm} 2]}$} 
\end{array}
$ & singularity type & MW group \\ \hline
no.1 & $E^2_8 A_2$ & 0 \\ 
no.2 &  $D_{16} A_2$ & $\Z_2$ \\
no.3 & $E_7 D_{10}$ & $\Z\oplus \Z_2$ \\
no.4 & $A_{17}$ & $\Z\oplus \Z_3$ \\ 
no.5 & $E^3_6$ & $\Z_3$ \\
no.6 & $D_7 A_{11}$ & $\Z_4$ \\ \hline
\end{tabular}
\caption{\label{tab[212]fibration}Elliptic fibrations of the attractive K3 surface $S_{[2 \hspace{1mm} 1 \hspace{1mm} 2]}$, singularity types of the fibrations, and the Mordell--Weil groups.}
\end{table}
\endgroup

\begingroup
\renewcommand{\arraystretch}{1.1}
\begin{table}[htb]
\centering
  \begin{tabular}{|c|c|c|} \hline
$
\begin{array}{c}
\mbox{Elliptic fibrations of}\\
\mbox{$S_{[2 \hspace{1mm} 0 \hspace{1mm} 2]}$} 
\end{array}
$ & singularity type & MW group \\ \hline
no.1 & $E^2_8 A^2_1$ & 0 \\ 
no.2 &  $E_8 D_{10}$ & 0 \\
no.3 & $D_{16} A^2_1$ & $\Z_2$ \\
no.4 & $E^2_7 D_4$ & $\Z_2$ \\ 
no.5 & $E_7 D_{10} A_1$ & $\Z_2$ \\
no.6 & $A_{17} A_1$ & $\Z_3$ \\ 
no.7 & $D_{18}$ & 0 \\ 
no.8 &  $D_{12} D_6$ & $\Z_2$ \\
no.9 & $D^2_8 A^2_1$ & $\Z_2 \oplus \Z_2$ \\
no.10 & $A_{15} A_3$ & $\Z_4$ \\ 
no.11 & $E_6 A_{11}$ & $\Z\oplus \Z_3$ \\
no.12 & $D^3_6$ & $\Z_2\oplus\Z_2$ \\ 
no.13 & $A^2_9$ & $\Z_5$ \\ \hline
\end{tabular}
\caption{\label{tab[202]fibration}Elliptic fibrations of the attractive K3 surface $S_{[2 \hspace{1mm} 0 \hspace{1mm} 2]}$, singularity types of the fibrations, and the Mordell--Weil groups.}
\end{table}
\endgroup

\newpage

\section{Fiber types of the extremal rational elliptic surfaces}
\label{secB}
We list the types of the singular fibers for the extremal rational elliptic surfaces \cite{MP} in Table \ref{tabextremalRES} below. The complex structure of an extremal rational elliptic surface is uniquely determined by the fiber type except the surface $X_{[0^*, \hspace{1mm} 0^*]}(j)$, and the complex structure of the extremal rational elliptic surface $X_{[0^*, \hspace{1mm} 0^*]}(j)$ depends on the j-invariant $j$ of the elliptic fiber \cite{MP}.

\begingroup
\renewcommand{\arraystretch}{1.5}
\begin{table}[htb]
\centering
  \begin{tabular}{|c|c|c|} \hline
$
\begin{array}{c}
\mbox{Extremal rational}\\
\mbox{elliptic surface} 
\end{array}
$ & fiber type & singularity type \\ \hline
$X_{[II, \hspace{1mm} II^*]}$ & $II^*$, $II$ & $E_8$  \\ \hline
$X_{[III, \hspace{1mm} III^*]}$ & $III^*$, $III$ & $E_7A_1$  \\ \hline
$X_{[IV, \hspace{1mm} IV^*]}$ & $IV^*$, $IV$ & $E_6A_2$ \\ \hline
$X_{[0^*, \hspace{1mm} 0^*]}(j)$ & $I_0^*$, $I_0^*$ & $D_4^2$ \\ \hline
$X_{[II^*, \hspace{1mm} 1, 1]}$ & $II^*$ $I_1$ $I_1$ & $E_8$ \\ \hline
$X_{[III^*, \hspace{1mm} 2, 1]}$ & $III^*$ $I_2$ $I_1$ & $E_7A_1$ \\ \hline
$X_{[IV^*, \hspace{1mm} 3, 1]}$ & $IV^*$ $I_3$ $I_1$ & $E_6A_2$ \\ \hline
$X_{[4^*, \hspace{1mm} 1, 1]}$ & $I_4^*$ $I_1$ $I_1$ & $D_8$ \\ \hline
$X_{[2^*, \hspace{1mm} 2, 2]}$ & $I_2^*$ $I_2$ $I_2$ & $D_6A_1^2$ \\ \hline
$X_{[1^*, \hspace{1mm} 4, 1]}$ & $I^*_1$ $I_4$ $I_1$ & $D_5A_3$ \\ \hline
$X_{[9, 1, 1, 1]}$ & $I_9$ $I_1$ $I_1$ $I_1$ & $A_8$ \\ \hline
$X_{[8, 2, 1, 1]}$ & $I_8$ $I_2$ $I_1$ $I_1$ & $A_7A_1$ \\ \hline 
$X_{[6, 3, 2, 1]}$ & $I_6$ $I_3$ $I_2$ $I_1$ & $A_5A_2A_1$ \\ \hline
$X_{[5, 5, 1, 1]}$ & $I_5$ $I_5$ $I_1$ $I_1$ & $A_4^2$  \\ \hline
$X_{[4, 4, 2, 2]}$ & $I_4$ $I_4$ $I_2$ $I_2$ & $A_3^2A_1^2$ \\ \hline
$X_{[3, 3, 3, 3]}$ & $I_3$ $I_3$ $I_3$ $I_3$ & $A_2^4$ \\ \hline
\end{tabular}
\caption{\label{tabextremalRES}Fiber types of the extremal rational elliptic surfaces are listed.}
\end{table}  
\endgroup

\end{document}